\documentclass[12pt]{article} 
\usepackage{times, amsmath, amsfonts, amsthm, physics, amssymb}

\usepackage[dvips]{graphicx,graphics}
\graphicspath{ {./graphs/} }
\usepackage{algorithm}
\usepackage{algpseudocode}
\usepackage{subcaption}
\usepackage{pdfpages}
\usepackage{blkarray}
\usepackage{bbm}
\usepackage{float}
\usepackage{bbm}
\usepackage[normalem]{ulem}
\usepackage{multirow}
\usepackage{gensymb}
\usepackage{textcomp}

\usepackage{color, colortbl}
\definecolor{Gray}{gray}{0.9}

\newtheorem{theorem}{Theorem}
\newtheorem{prop}{Proposition}

\DeclareMathSymbol{\sminus}{\mathbin}{AMSa}{"39}
\usepackage{url}
\usepackage{hyperref}
\hypersetup{colorlinks,%
citecolor=blue,%
filecolor=blue,%
linkcolor=red,%
urlcolor=blue,%
}

\DeclareMathOperator*{\argmin}{arg\,min}

\begin{document}

\title{Brain Networks Flow-Topology via Variance Minimization in the Wasserstein Space}
\author{Sixtus Dakurah\\
Department of Statistics, University of Wisconsin-Madison\\
\textit{sdakurah@wisc.edu}}

\date{}
\maketitle

\begin{abstract}
This work introduces a novel framework for testing topological variability in weighted networks by combining Hodge decomposition with Wasserstein variance minimization. Traditional approaches that analyze raw edge weights are susceptible to noise driven perturbations, limiting their ability to detect meaningful structural differences between network populations. Network signals are decomposed into various components using combinatorial Hodge theory, then topological disparity is quantified via the $2$-Wasserstein distance between persistence diagrams. The test statistic measures variance reduction when comparing within group to between group dispersions in the Wasserstein space. Simulations demonstrate that the proposed method suppresses small random perturbations while maintaining sensitivity to genuine topological differences, particularly when applied to Hodge decomposed flows rather than raw edge weights. The framework is applied to functional brain networks from the Multimodal Treatment of ADHD dataset, comparing cannabis users and non-users.\\


\end{abstract}

\section{Introduction}

Network analysis has become indispensable for understanding complex systems across scientific domains, from social networks to biological systems. In neuroscience, the brain's functional organization manifests as a complex network of interconnected regions whose coordinated activity underlies cognitive processes \cite{bullmore2009complex,hagmann2008mapping}. Functional magnetic resonance imaging (fMRI) enables non-invasive measurement of these functional brain networks through statistical dependencies between regional activity patterns, producing weighted connectivity graphs where nodes represent brain regions and edges encode correlation strengths. A fundamental challenge in network neuroscience is determining whether observed differences between network populations, for example, healthy controls versus clinical groups, or different experimental conditions reflect genuine structural variation or merely noise-driven perturbations. This distinction is critical for valid statistical inference and biological interpretation.

Traditional network comparison methods typically operate directly on edge weights or derived graph-theoretic measures such as clustering coefficients, path lengths, and modularity \cite{dakurah2025topologically,newman2018networks}. While these approaches have yielded important insights into network organization, they often lack robustness to noise: small random fluctuations in edge weights can substantially alter graph metrics, leading to spurious detection of group differences. Moreover, conventional permutation tests applied to raw connectivity matrices do not account for the topological structure of networks, potentially generating null distributions that distort the intrinsic organizational features of the data \cite{chung2019rapid,dakurah2025discrete,dakurah2025topologically,nichols2002nonparametric}. This sensitivity to noise is particularly problematic in neuroimaging, where measurement variability from scanner noise, physiological artifacts, and individual differences creates substantial edge-level fluctuations even when underlying network architecture remains stable.

Recent advances in topological data analysis offer promising alternatives for robust network comparison. Persistent homology, a core technique from algebraic topology, characterizes multi-scale structural features by tracking the birth and death of connected components and cycles across network filtrations \cite{edelsbrunner2008persistent,zomorodian2005computing}. By representing networks through persistence diagrams, persistent homology compresses complex connectivity patterns into topological signatures that are invariant to continuous deformations. These representations have proven valuable for characterizing network structure in diverse applications \cite{anand2021hodge,dakurah2022modelling,dakurahregistration,songdechakraiwut2020topological}, offering robustness properties that complement traditional graph metrics. However, persistence diagrams alone do not address how to decompose edge-level signals into interpretable components or how to construct statistical tests that properly account for topological variability.

Complementing the global perspective of persistent homology, combinatorial Hodge theory provides a framework for decomposing network signals into orthogonal flow components with distinct structural interpretations \cite{lim2020hodge,schaub2021signal}. The Helmholtz-Hodge decomposition, henceforth referred to as Hodge decomposition, originally developed in differential geometry and extended to discrete networks, separates edge flows into gradient flows (representing potential-driven or non-loop processes), curl flows (capturing local loop circulation), and harmonic flows (encoding global loop structures) \cite{barbarossa2020topological,schaub2018flow}. In this formulation, gradient flows constitute non-loop flows, while curl and harmonic flows together form the loop-flow components, representing local and global circulations respectively. This decomposition isolates functionally distinct aspects of network organization. Recent work has demonstrated that Hodge decomposition can reveal nuanced patterns in brain connectivity that are obscured when analyzing raw edge weights \cite{anand2024hodge,dakurah2025discrete,dakurahregistration}.

The Wasserstein distance, originating in optimal transport theory, provides a natural metric for comparing probability distributions and has recently gained traction in network analysis through its application to persistence diagrams \cite{agueh2011barycenters,rabin2011wasserstein}. Unlike pointwise comparison metrics, the Wasserstein distance accounts for the geometry of the space and admits a meaningful notion of average (the Wasserstein barycenter) in the space of distributions. For networks represented through persistence diagrams, the 2-Wasserstein distance has a closed-form expression when birth and death values are treated as one-dimensional distributions, enabling efficient computation without requiring optimal transport solvers \cite{songdechakraiwut2020topological}. The Wasserstein variance, defined as the minimal average squared distance from distributions to their barycenter, quantifies dispersion in topological feature spaces and provides a principled measure of variability within and between network populations.

Despite these methodological advances, existing approaches for network comparison face several limitations. First, most methods operate on raw edge weights without decomposing networks into structurally interpretable components, potentially conflating different sources of variation. Second, standard statistical tests on persistence diagrams do not account for how topological features are derived from underlying connectivity patterns, limiting their ability to localize specific edges driving group differences. Third, few methods explicitly address the challenge of suppressing noise-driven variance while preserving genuine topological differences. Recent work has begun exploring variance minimization in Wasserstein spaces for network analysis \cite{dakurah2025discrete,dakurah2025robust,dakurah2024subsequence,dakurah2022modelling,dakurah2025maxtda}, but the integration of Hodge decomposition with topological variance minimization remains largely unexplored.

This work introduces a unified framework for testing topological variability in weighted networks by integrating Hodge decomposition with Wasserstein variance minimization. The method mitigates the influence of noise by first decomposing Edge Flows into Loop Flows and Non-Loop Flows, then quantifying topological disparity through persistence diagrams. The main insight is that each flow component responds differently to perturbations: Non-Loop Flows (gradient flows) capture potential-driven directional processes that tend to be robust to local noise, while Loop Flows (curl and harmonic flows) reflect local and global circulation patterns that are more sensitive to structural or connectivity changes. By applying Wasserstein variance minimization separately to each component, the framework identifies which aspects of network topology contribute most to population-level differences while controlling for noise-induced variability.

The proposed test statistic measures the difference between within-group topological variance (networks compared to their group barycenter) and between-group variance (all networks compared to individual group barycenters). Under the null hypothesis of topological equivalence, this statistic should be small, as networks from different groups exhibit similar dispersion patterns. Under the alternative hypothesis of genuine topological differences, the statistic increases, reflecting reduced within-group variance relative to between-group comparisons. This formulation naturally extends analysis of variance principles to the Wasserstein space while respecting the geometric structure of persistence diagrams \cite{hu2016matched,ortega2018graph}.

Systematic simulations and application to real neuroimaging data are used to validate the proposed framework. Simulations show that Wasserstein variance minimization effectively suppresses noise-driven perturbations when applied to Hodge-decomposed flows, particularly the gradient (non-loop) component. Networks differing only by random perturbation magnitude yield high $p$-values, indicating successful noise filtering—when compared through Non-Loop Flows, whereas raw edge-based comparisons produce false positives at larger sample sizes. These findings in this work confirm that decomposition prior to comparison provides greater robustness than analyses based on aggregate connectivity, and makes several methodological contributions. First, we develop a test statistic based on Wasserstein variance minimization that is theoretically justified under the null hypothesis and provides a clear interpretation: small values indicate topological equivalence. Second, we establish that Hodge decomposition isolates network structure into components with differential noise sensitivity, enabling targeted analysis of specific architectural features. Third, we provide a localization procedure for identifying specific edges contributing to group differences while controlling family-wise error rates. Finally, we demonstrate through comprehensive simulations that the proposed approach outperforms raw edge-based methods in distinguishing genuine topological differences from noise.

The remainder of this paper is organized as follows. Section~\ref{sec:methods} develops the mathematical framework, including simplicial complexes, graph filtrations, Hodge decomposition, Wasserstein distance and variance, and the variance minimization test statistic. Section~\ref{sec:exp-results} presents simulation studies validating the method's noise suppression properties and establishes appropriate behavior under null and alternative hypotheses. Section~\ref{sec:app} applies the framework to functional brain networks from the ACPI-MTA dataset, comparing cannabis users and non-users across Edge, Loop, and Non-Loop Flow components. Section~\ref{sec:disc} discusses theoretical implications, practical advantages, biological interpretation, limitations, and future directions for topological network analysis.

\section{Methods}
\label{sec:methods}

\subsection{Graphs as Simplicial Complexes}
A $k$-simplex $\sigma_k = (v_0, \cdots, v_k)$ is a $k$-dimensional polytope of nodes $v_0, \cdots, v_k$.
A simplicial complex $K$ is a finite set of simplices such that for any $\sigma_k^i, \sigma_k^j \in K$, $\sigma_k^i \cap \sigma_k^j$ is a face of both simplices; and a face of any $\sigma_k^i \in K$ is also a simplex in $K$ \cite{edelsbrunner2008persistent}. 
A 0-skeleton is a simplicial complex consisting of only nodes, whiles a 1-skeleton consists of nodes and edges.
A $k$-chain is a finite sum of simplices. For two successive chain groups $\mathcal{K}_k$ and $\mathcal{K}_{k-1}$, the boundary operator $\partial_k: \mathcal{K}_k \longrightarrow \mathcal{K}_{k-1}$ for each $\sigma_k$ is given by
$$
    \partial_k(\sigma_k) = \sum_{i=0}^k(-1)^i(v_0, \cdots, \widehat{v_i}, \cdots, v_k)
    \label{eqn:boundary_map},
$$
where $(v_0, \cdots, \widehat{v_i}, \cdots, v_k)$ gives the $k$-$1$ faces of $\sigma_k$ obtained by deleting node $\widehat{v_i}$. The matrix representation $\mathbb{B}_k = (\mathbb{B}_k^{ij})$ of the boundary operator is given by
\begin{equation*}
	\mathbb{B}_k^{ij} =
	\begin{cases}
		1, 	& \text{if } \sigma^i_{k-1}  \subset \sigma^j_{k}  ~~\text{and}~~ \sigma^i_{k-1} \sim \sigma^j_{k}\\
		-1, & \text{if } \sigma^i_{k-1}  \subset \sigma^j_{k}  ~~\text{and}~~ \sigma^i_{k-1} \nsim \sigma^j_{k}\\
		0,  & \text{if } \sigma^i_{k-1}  \not \subset \sigma^j_{k}
	\end{cases},
	\label{eq:boundarymatrix}
\end{equation*}
where $\sim$ and $\nsim$ denote similar and dissimilar orientations.

\begin{figure}[ht]
 \centering
 \includegraphics[width =\linewidth]{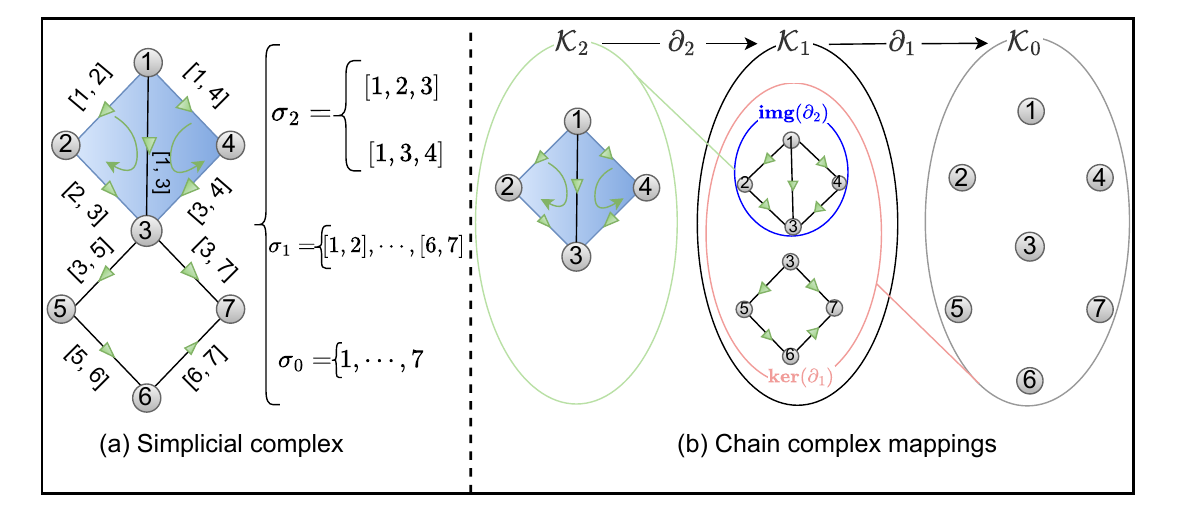}
 \caption{(a) A 2-dimensional simplicial complex. (b) A chain mapping from higher to lower dimensional simplices.}
 \label{fig:schematic-boundary-operators}
\end{figure}

The kernel of $\partial_k$ is denoted as $\mathcal{Z}_k = ker(\partial_k)$ and its image denoted as $\mathcal{B}_{k} = im(\partial_{k+1})$. The elements of $\mathcal{Z}_k$ and $\mathcal{B}_{k}$  are known as $k$-cycles and $k$-boundaries respectively \cite{edelsbrunner2008persistent,lim2020hodge}. From the foregoing, graphs are 1-skeletons, and simplicial complexes generalizes graphs to higher-dimensions. Figure \ref{fig:schematic-boundary-operators} illustrates the computation of the boundary matrices and corresponding kernels.

\subsubsection{Birth-Death Decomposition of Graphs}
We consider a graph $\mathcal{X} = (V, E, {w})$ with node set $V$, edge set $E$, and edge weight matrix ${w} = (w_{ij})$.
The binary graph $\mathcal{X}_{\epsilon} = (V, E, w_\epsilon)$ is a graph with the binary edge weights $w_\epsilon = (w_{\epsilon, ij})$ where $w_{\epsilon, ij}=1$ if $w_{ij} > \epsilon$ and is $0$ otherwise. 
A graph filtration is a collection of a sequence of nested binary networks \cite{chung2019exact}
$
\mathcal{X}_{{\epsilon}_0} \supset ... \supset \mathcal{X}_{{\epsilon}_k},
$
where ${{\epsilon}_0} < ... < {{\epsilon}_k}$ are ordered thresholds.
As $\epsilon$ increases, more edges are disconnected, increasing the number of connected components $(\beta_0)$, and decreasing the number of cycles $(\beta_1)$. It has been shown that $\beta_0$ and $\beta_1$ are monotone over filtration \cite{chung2019exact}.
The filtration tracks the persistence of connected components and $1$-cycles \cite{zomorodian2005computing}. The persistence of a connected component or 1-cycle that appears at filtration value $b_i$ and disappears at filtration value $d_i$ is given by the interval $\left[b_i,  d_i\right]$. A finite collection of $\left[b_i,  d_i\right]$ can be summarized in the form of a \textit{barcode}.

 During the filtration,  $0$-cycles are born at $b_i$ and never dies ($\infty$) whiles $1$-cycles are born when the graph is formed ($-\infty$) and die at $d_i$. Ignoring the infinite values in both cases, we can represent the birth $B(\mathcal{X})$ and death $D(\mathcal{X})$ values as barcodes:
$$ 
B(\mathcal{X}) = b_1  < b_2 < \dots < b_{p}, \hspace{0.1cm}  D(\mathcal{X}) = d_1 < d_2 < \dots < d_{q}.
$$
It follows that $B(\mathcal{X}) \cup D(\mathcal{X}) = (w_{ij})$ and $B(\mathcal{X}) \cap D(\mathcal{X}) = \emptyset$ \cite{songdechakraiwut2020topological}. This decomposition will be applied to the flow-topology of graphs to study its variability in the Wasserstein space.

\subsection{Flow-Topology of Simplicial Complexes}
\label{sec:hodge_decompose}
We use the combinatorial Hodge theory to decompose signals on a simplicial complex into it's globally and locally circular components that characterize information flow.\\

\subsubsection{Hodge Decomposition}
\begin{figure}[ht!]
\centering
\includegraphics[width=1\linewidth,clip=true]{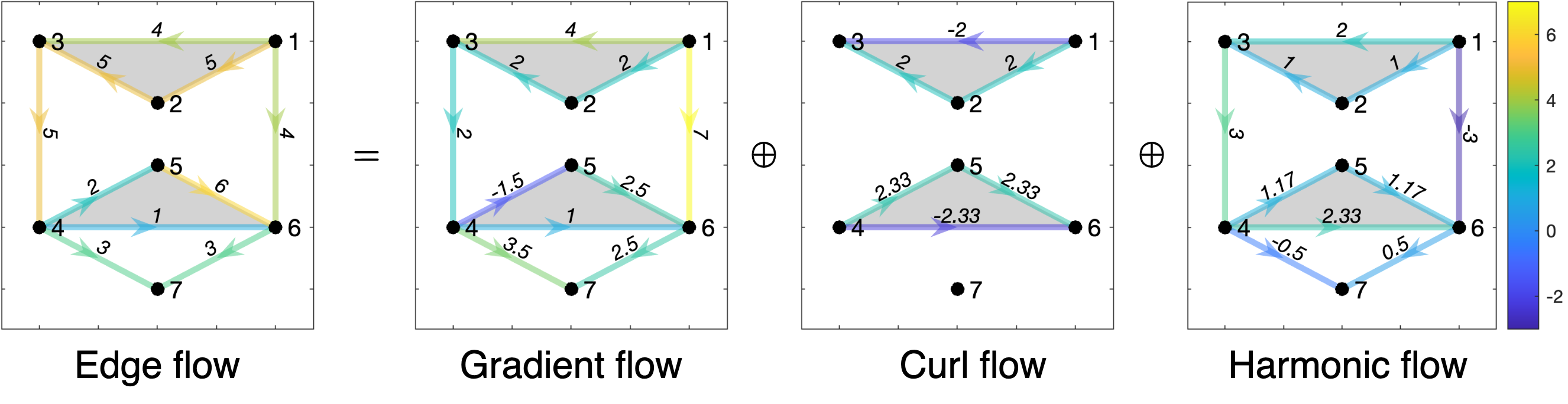}
\caption{Hodge decomposition of the edge flow into 3 orthogonal subspaces. Curl flow identifies the triangles (filled-in).}
\label{fig:hdcmp}
\end{figure}

The boundary matrix can be used to construct a higher-dimensional generalization of the graph Laplacian, known as the Hodge Laplacian \cite{lim2020hodge}:
$$
\mathbb{L}_1 = \mathbb{B}_{2}\mathbb{B}_{2}^\top + \mathbb{B}_{1}^\top\mathbb{B}_{1}.
$$
$\mathbb{L}_1$ acts as a shift operator on signals defined on 1-simplices. 
The Hodge Laplacian is positive semi-definite, hence its eigenvalues can be viewed as non-negative frequencies, whiles the corresponding eigenvectors represent a specific type of signal-smoothness \cite{lim2020hodge,schaub2021signal}. This signal smoothness can be formalized through the {\em Hodge decomposition} theorem which states that signals in a $1$-simplex admits a decomposition \cite{lim2020hodge}:
$$
\mathbb{R}^{|E|} = im(\mathbb{B}_{k+1}) \oplus im(\mathbb{B}_{k}^\top) \oplus ker(\mathbb{L}_k).
$$
The space $im(\mathbb{B}_1^\top) = \{ \mathbf{g}= \mathbb{B}_1^\top \mathbf{v}, \mathbf{v} \in \mathbb{R}^{|V|} \}$ is the space of gradient flows. The space $im(\mathbb{B}_2) = \{\mathbf{c} = \mathbb{B}_2 \mathbf{t}\}$ where $\mathbf{t}$ is a vector of $2$-simplex potentials is the curl flow space. Let $\mathbf{e}$ denote the edge flow. Using orthogonality of the decomposition, we can numerically solve for the vectors $\mathbf{g}$ and $\mathbf{c}$:
$$
\min_{\mathbf{v}}|| \mathbb{B}_1^\top \mathbf{v} - \mathbf{e} ||_2, \quad \min_{\mathbf{t}}|| \mathbb{B}_2 \mathbf{t} - \mathbf{e} ||_2.
$$
The harmonic space $ker(\mathbb{L}_1)$ which characterizes global circulation is computed by subtraction since $\mathbf{e} = \mathbf{g} \oplus \mathbf{c} \oplus \mathbf{h}$, where $\mathbf{h} \in ker(\mathbb{L}_1)$. The curl and harmonic spaces are circular flows and illuminates feedback loops built into a network.
Figure \ref{fig:hdcmp} shows such a decomposition.
The local and global circular segregation of information flow through the decomposition reveals nuisances in the variation of networks that might not be apparent when considering only the edge flow. Functional brain networks which are densely connected can be efficiently studied using this decomposition in the Wasserstein space by observing topological variability.

\subsection{Variance Minimization in the Wasserstein Space} The Wasserstein distance has emerged as a popular metric for quantifying topological disparity between networks \cite{songdechakraiwut2020topological,anand2021hodge}.\\

\subsubsection{Wasserstein Distance between Graphs}
Let $\mathcal{X}^1$ and $\mathcal{X}^2$ be networks with corresponding persistence diagrams $P_{\mathcal{X}^1}$ and $P_{\mathcal{X}^2}$. The elements of $P_{\mathcal{X}^1}$ and $P_{\mathcal{X}^2}$ are scatter points and can be modelled after the Dirac delta functions. The r-Wasserstein distance between $P_{\mathcal{X}^1}$ and $P_{\mathcal{X}^2}$ is defined as
$$
\mathfrak{L}_r(P_{\mathcal{X}^1}, P_{\mathcal{X}^2}) = \inf_{\tau: P_{\mathcal{X}^1}\xrightarrow{} P_{\mathcal{X}^2}} \left( \sum_{x \in P_{\mathcal{X}^1}} ||x - \tau(x)||^r \right)^{\frac{1}{r}},
$$
where $\tau$ is the bijection between $P_{\mathcal{X}^1}$ and  $P_{\mathcal{X}^2}$. In this work, we only consider $r = 2$. Since $P_{\mathcal{X}^1}$ and $P_{\mathcal{X}^2}$ are one dimensional, $\mathfrak{L}_2$ admits a close form summarized in this theorem.
\begin{theorem}
The 2-Wasserstein distance between the persistence diagrams admits the close form
$$
\mathfrak{L}_2(P_{\mathcal{X}^1}, P_{\mathcal{X}^2}) = \left(\sum_{i = 1}^p |b_i^1 - b_i^2|^2 \right)^{\frac{1}{2}}
+ 
\left(\sum_{i = 1}^q |d_i^1 - d_i^2|^2 \right)^{\frac{1}{2}},
$$
where $b_i^{k}$ and $d_i^{k}$ is the $i$-th smallest birth and death values.
\end{theorem}

\subsubsection{Center in the Wasserstein Space}
One useful property of the Wasserstein metric is the concept of center (average) defined in the space of probability measures.
Let ${\{\mathcal{P}_j\}}_{j \in J} \subset \mathbb{R}^n$ be a family of point clouds, and assume these point clouds to be discrete density distributions. The barycenter is a weighted average point cloud ${\mathcal{P}^*}$, defined through the minimizer
$$
{\zeta}(w_j, \mathcal{P}_j)_{j \in J} \in \argmin_{{\mathcal{P}^*}} E ({\mathcal{P}^*}) = \sum_{j\in J} w_j \mathfrak{L}_2(\mathcal{P}_j, {\mathcal{P}^*}),
$$
where $w_j \ge 0$ and is constrained such that $\sum_{j \in J} w_j = 1$ \cite{agueh2011barycenters,rabin2011wasserstein}.
In the special case where ${\{\mathcal{P}_j\}}_{j \in J} \subset \mathbb{R}^n$, $n = 1$, the barycenter can be computed by taking the average of sorted point clouds \cite{agueh2011barycenters,rabin2011wasserstein}. For a graph network $\mathcal{X}$, $\mathcal{P}_j = P_{\mathcal{X}}$. If we let $\tau(x_i)$ to be the permutation of the $i$-th element in $\mathcal{P}_j$ such that $\tau(x_i) \le \tau(x_{i+1})$, then the barycenter reads

$$
\left({\zeta}(w_j, \mathcal{P}_j)_{j \in J} \right)_i = \sum_{j\in J} w_j \tau(x_{i}), \quad  1 \le i \le m.
$$
This barycenter can be computed in $O(m \log (m) )$.\\

\subsubsection{Wasserstein Variance}
From the Wasserstein barycenter, we can define the Wasserstein variance as the resulting minimal value. More formally, we define the Wasserstein variance for a set of point clouds $\{\mathcal{P}_j\}_{j \in J} \subset \mathbb{R}^n$ and weights $(w_j)$ as 
$$
 \inf_{\mathcal{P} \subset \mathbb{R}^n} \left( \sum_{j \in J}w_j \mathfrak{L}_2(\mathcal{P}_j, {\mathcal{P}})\right) = \sum_{j \in J}w_j \mathfrak{L}_2(\mathcal{P}_j, {\mathcal{P}^*}).
$$
We denote the variance by WV, and to emphasize the point clouds $\{\mathcal{P}_j\}$ from which it is computed, we index it as $$WV(\{\mathcal{P}_j\}, \mathcal{P}^*) = \sum_{j \in J}w_j \mathfrak{L}_2(\mathcal{P}_j, {\mathcal{P}^*}).$$
The Wasserstein variance is a practical tool to quantify the topological variability of different networks. A low variance implies similar distributions. We develop our testing framework around this realization. 

\subsection{Framework for Testing Topological Variability}
We now develop the framework for testing for topological variability between groups of networks.\\

\subsubsection{Global Test of Network Differences}
Consider two groups of networks $\mathcal{N}_1 = \{\mathcal{X}^1, \cdots, \mathcal{X}^{m_1}\}$, $\mathcal{N}_2 = \{\mathcal{Y}^{1}, \cdots, \mathcal{Y}^{ m_2}\}$, and let $m = m_1 + m_2$. Denote the barycenters in the two groups by $\zeta_1$ and $\zeta_2$, computed from the persistence diagrams $\{\mathcal{P}_j^1\} = \{P_{\mathcal{X}^j}\}$, and  $\{\mathcal{P}_j^2\} = \{P_{\mathcal{Y}^j}\}$ respectively. We want to test the hypothesis
$$
H_0: \zeta_1 = \zeta_2,
$$
i.e., the two groups have the same barycenter, as opposed to the alternative hypothesis that their barycenters differ. Denote the collection of the two groups of point clouds as $\{\mathcal{P}_j\} = \{\{\mathcal{P}_j^1\}, \{\mathcal{P}_j^2\}\}$. We propose the following statistic $\mathcal{T}(\mathcal{N}_1, \mathcal{N}_2)$ for testing topological variability:
\begin{equation*}
   \mathcal{T} = \sum_{k = 1}^2 \left[ \frac{m_k}{m_k -1} WV(\{\mathcal{P}_j^k\}, \zeta_k) - \frac{m}{m-1}WV(\{\mathcal{P}_j\}, \zeta_k)\right]. 
\end{equation*}
This statistic measures the difference in the topological variability between networks in a specific group to it's barycenter and networks in all groups to the barycenter of a specific group. Let $\mathcal{T}_{H_0}$ and $\mathcal{T}_{H_1}$ denote the test statistic under the null and alternative hypothesis respectively, the following holds:
\begin{prop}
Small values of $\mathcal{T}$ provides evidence against the null hypothesis, further we have that $\mathcal{T}_{H_1} \le \mathcal{T}_{H_0}$. 
\label{prop:initial-test}
\end{prop}
This proposition asserts we can test for topological variability by minimizing the inherent variance in $\mathcal{T}$.\\

\subsubsection{Localization of Network Differences}
If $H_0$ in the above testing framework is rejected, it gives little information as to the specific connections that are contributing to this difference. Let $\mathcal{S}(E)$ be an index set identifying ordered connections for which $\mathcal{T}$ is minimum. Let $H_{0, k}$ denote the hypothesis involving the $k$-th ordered connection across the groups. We identify these connections as follows:
$$
\mathcal{S}(E) = \left\{ k: H_{0, k} \hspace{0.15cm} \text{is} \hspace{0.15cm} \text{false}, \forall k \in \{1, \cdots, |E| \} \right\}.
$$
To control the family-wise error rate, we applied infimum test statistic correction procedure \cite{chung2019exact}. 
\begin{theorem}
If $H_0$ is false, then $\exists \tilde{E} = \argmin_{\substack{{\tilde{E} \subset E}}} \mathcal{T}_{\tilde{E}}$, and as such $\mathcal{S}(E) \ne \varnothing$.
\label{them:initial-test}
\end{theorem}
The statistical significance of the tests in Proposition~\ref{prop:initial-test} and Theorem~\ref{them:initial-test} are inferred based on the non-parametric permutation testing procedure \cite{chung2019exact}.

\section{Experimental Results}
\label{sec:exp-results}

Since there is no topological ground truth in real data, we demonstrate the discriminatory power of the proposed Wasserstein variance minimization method through a series of simulations. The simulation is structured to distinguish between two identical networks but with minor variation in connectivity strengths. 

\subsection{Simulation Data and Experimental Setup}
\begin{figure}[ht!]
     \centering
     \begin{subfigure}[t]{0.325\linewidth}
         \centering
         \includegraphics[width=\textwidth]{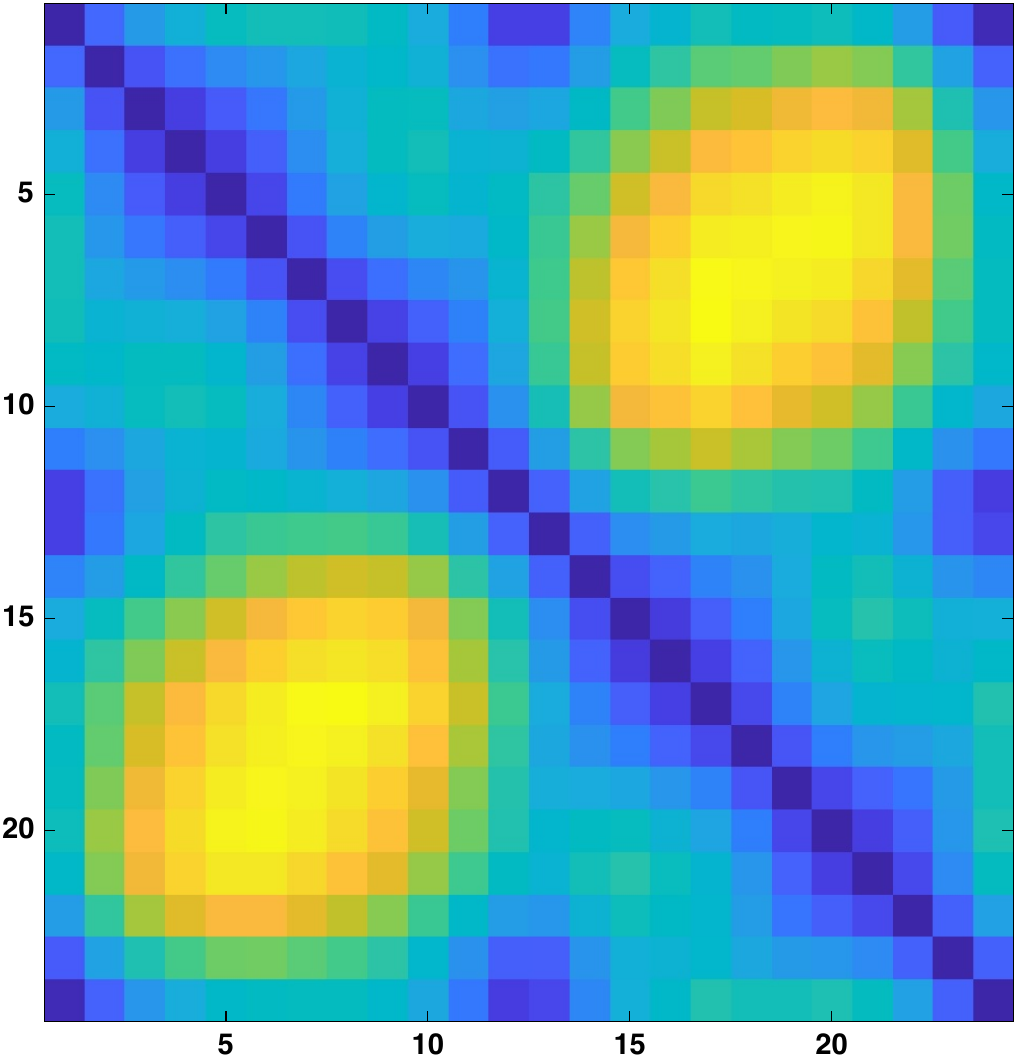}
         \caption{Base Group 1}
         \label{fig:b1}
     \end{subfigure}
     \hfill
     \begin{subfigure}[t]{0.325\linewidth}
         \centering
         \includegraphics[width=\textwidth]{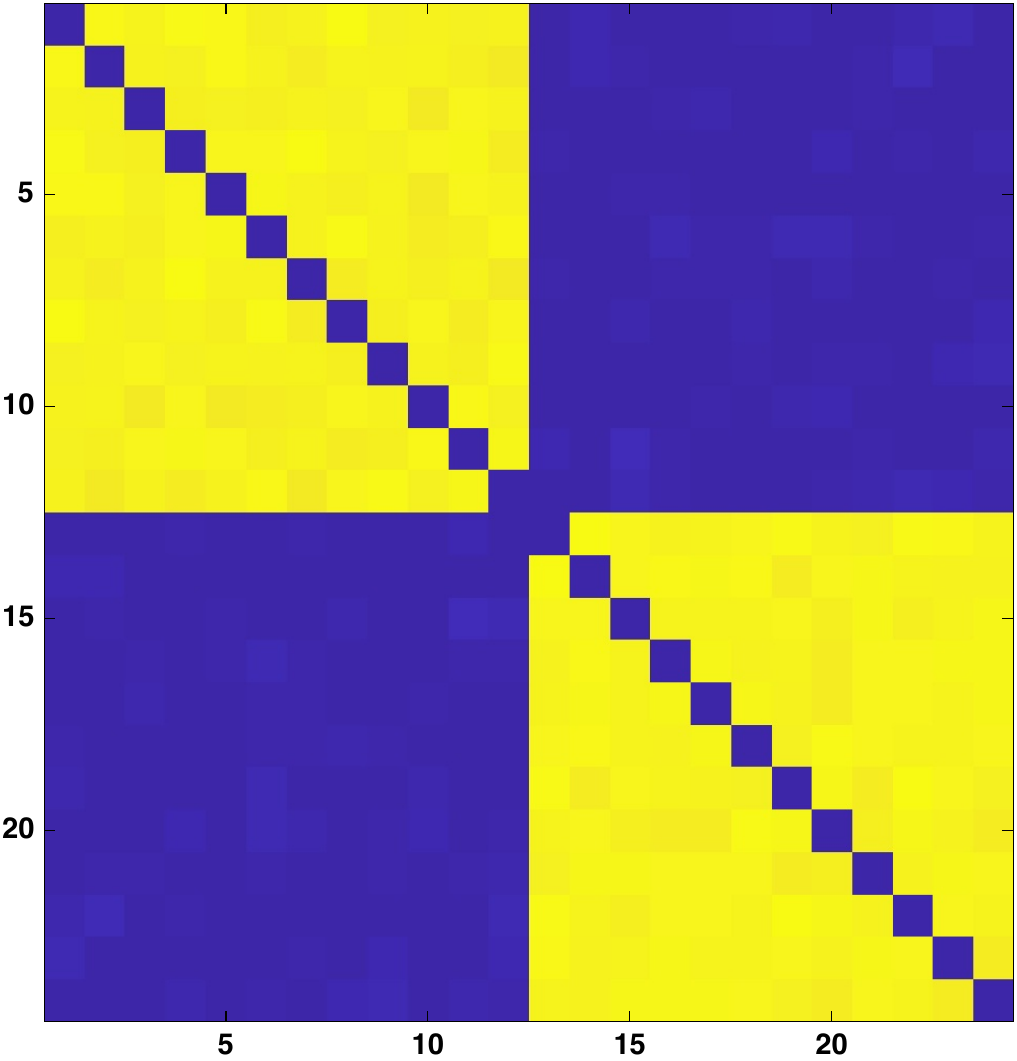}
         \caption{Modular Group 1}
         \label{fig:m1}
     \end{subfigure}
     \hfill
     \begin{subfigure}[t]{0.325\linewidth}
         \centering
         \includegraphics[width=\textwidth]{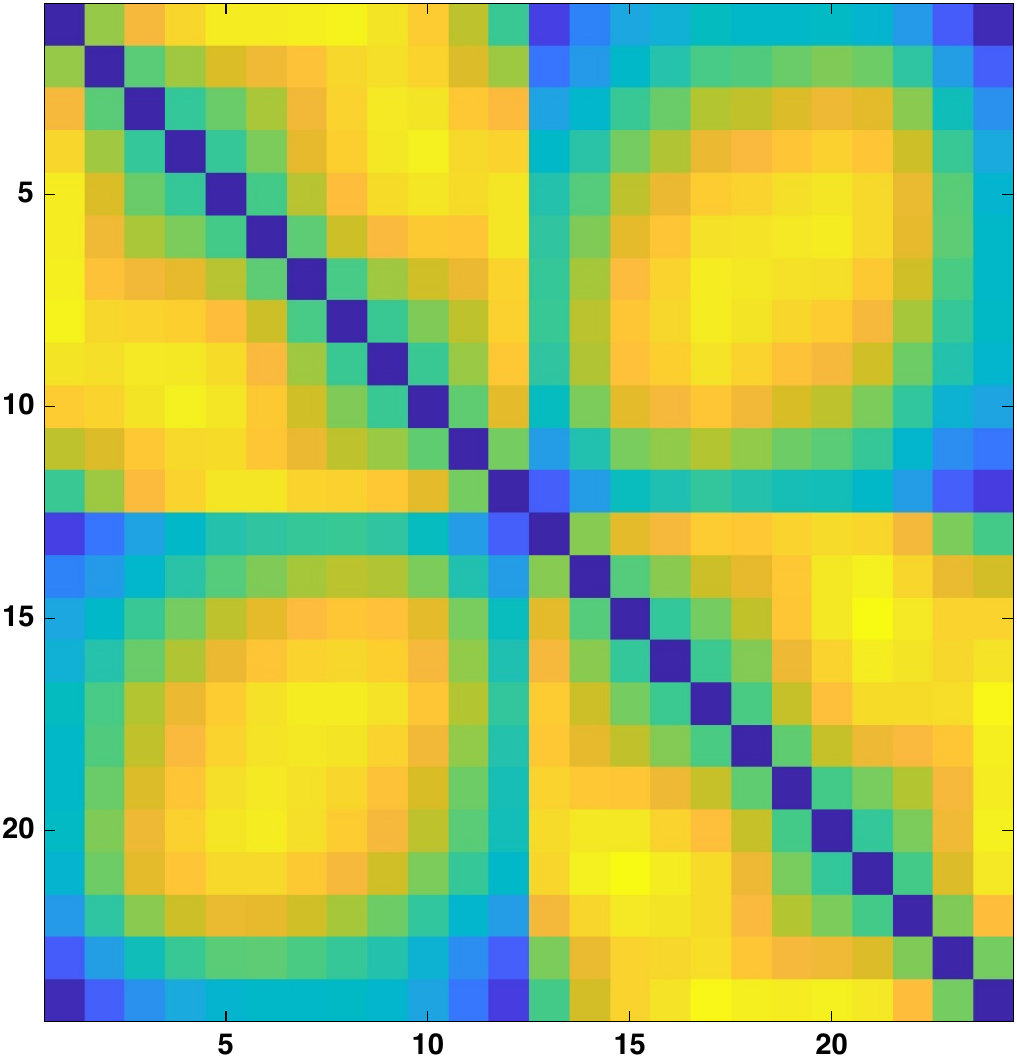}
         \caption{Base + Modular}
         \label{fig:bm1}
     \end{subfigure}
     \centering
     \begin{subfigure}[t]{0.325\linewidth}
         \centering
         \includegraphics[width=\textwidth]{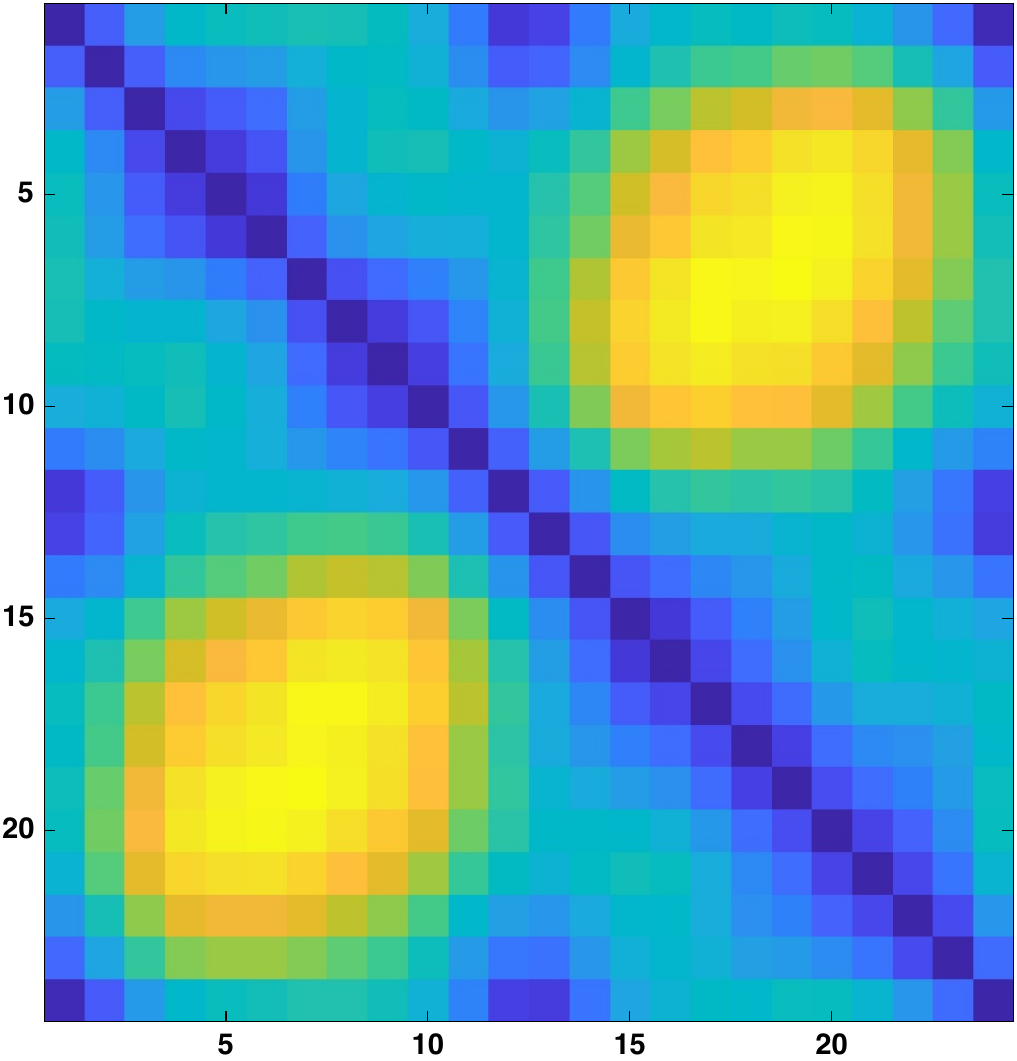}
         \caption{Base Group 2}
         \label{fig:b2}
     \end{subfigure}
     \hfill
     \begin{subfigure}[t]{0.325\linewidth}
         \centering
         \includegraphics[width=\textwidth]{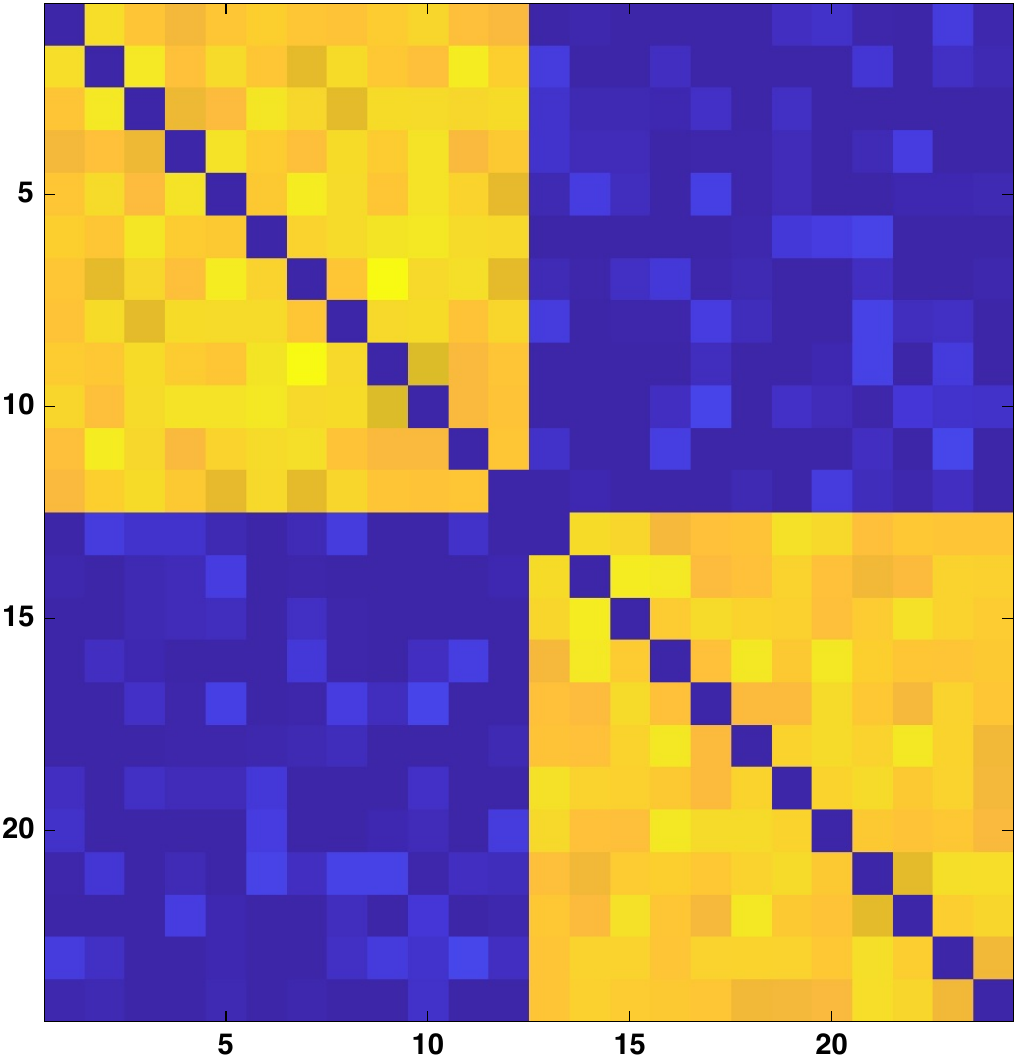}
         \caption{Modular Group 2}
         \label{fig:m2}
     \end{subfigure}
     \hfill
     \begin{subfigure}[t]{0.325\linewidth}
         \centering
         \includegraphics[width=\textwidth]{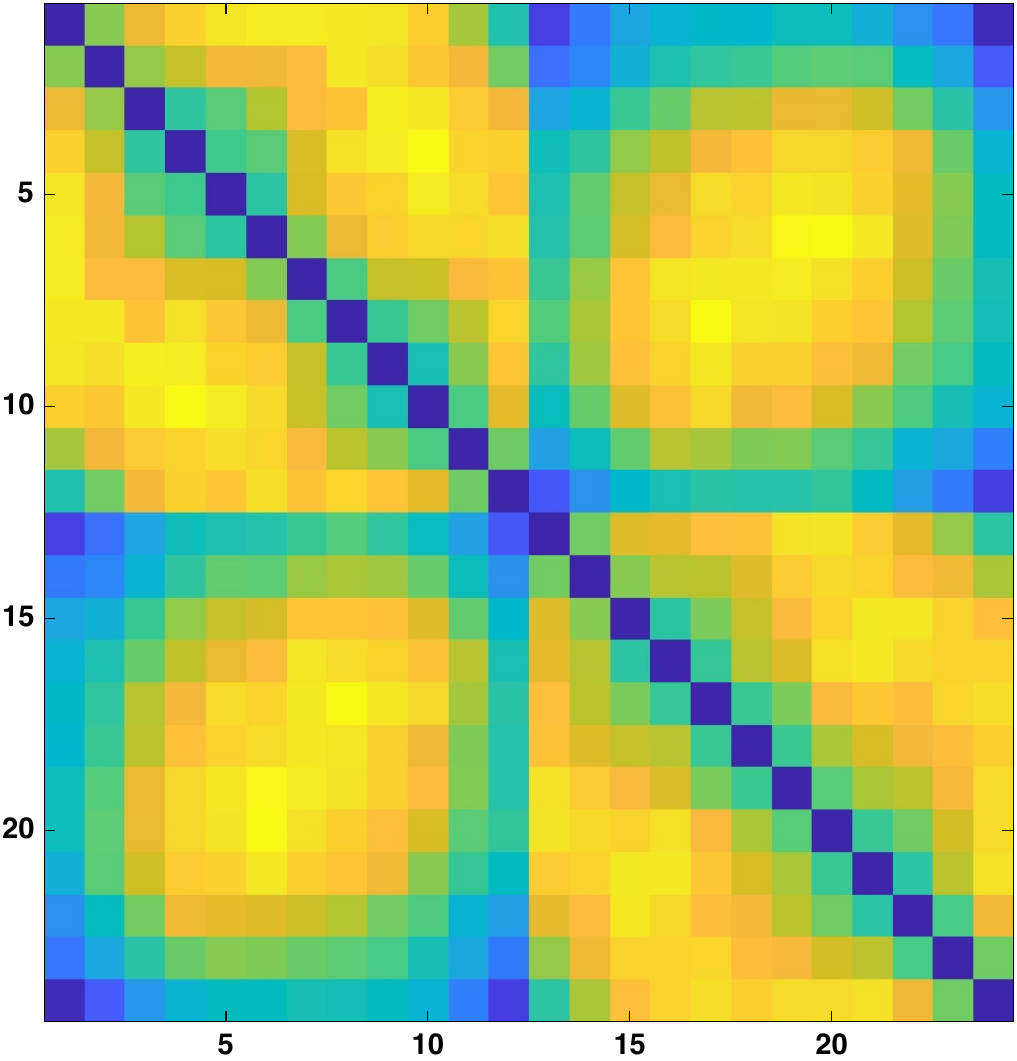}
         \caption{Base + Modular}
         \label{fig:bm2}
     \end{subfigure}
\caption{The two groups of networks to be compared. Top: (a) the based network comprising of two loops, (b) modular network with variation set at 0.2, (c) the combined base and modular network. Bottom: (d)  (d) the based network comprising of two loops, (e) modular network with variation set at 0.225, (e) the combined base and modular network. The color intensity is an indicator of the strength of the edge connectivity.}
    \label{fig:sn_dv_two_groups}
\end{figure}

We consider undirected weighted networks on $n=24$ nodes. Let $m=\binom{n}{2}$ denote the number of edges. To generate the data for the first group, we first construct a structured base network $A^{(1)}_{\text{base}} \in \mathbb{R}^{n\times n}$ containing two loops. We then add a modular perturbation $M^{(2)}$ with random variation drawn from a Gaussian distribution with standard deviation $0.2$. The sample level adjacency for the first and second groups are:
\begin{equation*}
A^{(1)}_i \;=\; A^{(1)}_{\text{base},i} \;+\; M^{(1)}_i, \quad A^{(2)}_i \;=\; A^{(2)}_{\text{base},i} \;+\; M^{(2)}_i,
\end{equation*}
where $M^{(1)}_i, M^{(2)}_i$ are modular perturbations with random variation drawn from a Gaussian distribution with standard deviation $0.225$. This yields two groups whose only difference is the strength of the modular perturbation. Figure~\ref{fig:sn_dv_two_groups} shows the two groups of networks. 

\begin{figure}[ht!]
     \centering
     \begin{subfigure}[t]{0.45\linewidth}
         \centering
         \includegraphics[width=\textwidth]{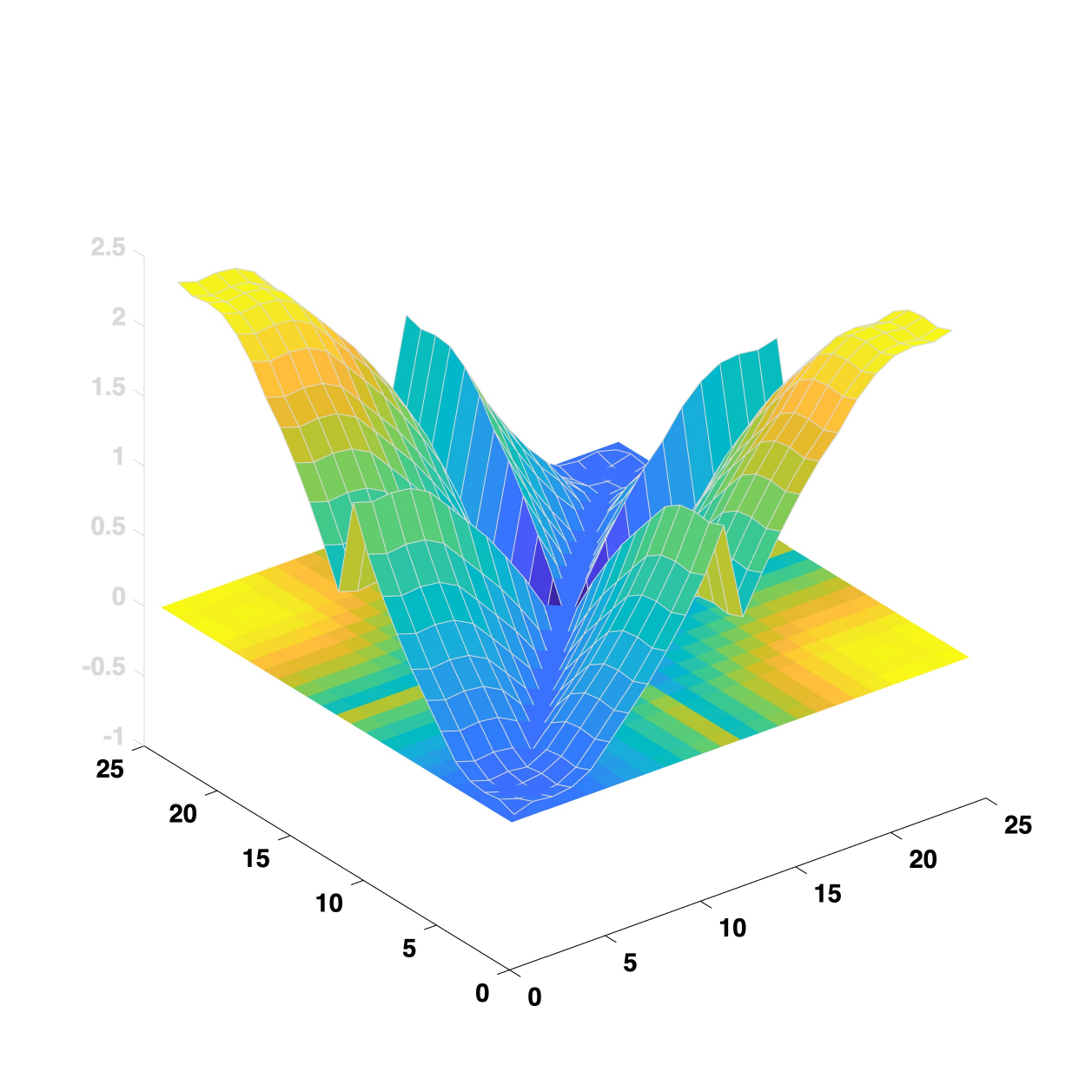}
         \caption{Non-Loop Flow - Group 1}
         \label{fig:sg1}
     \end{subfigure}
     \hfill
     \begin{subfigure}[t]{0.45\linewidth}
         \centering
         \includegraphics[width=\textwidth]{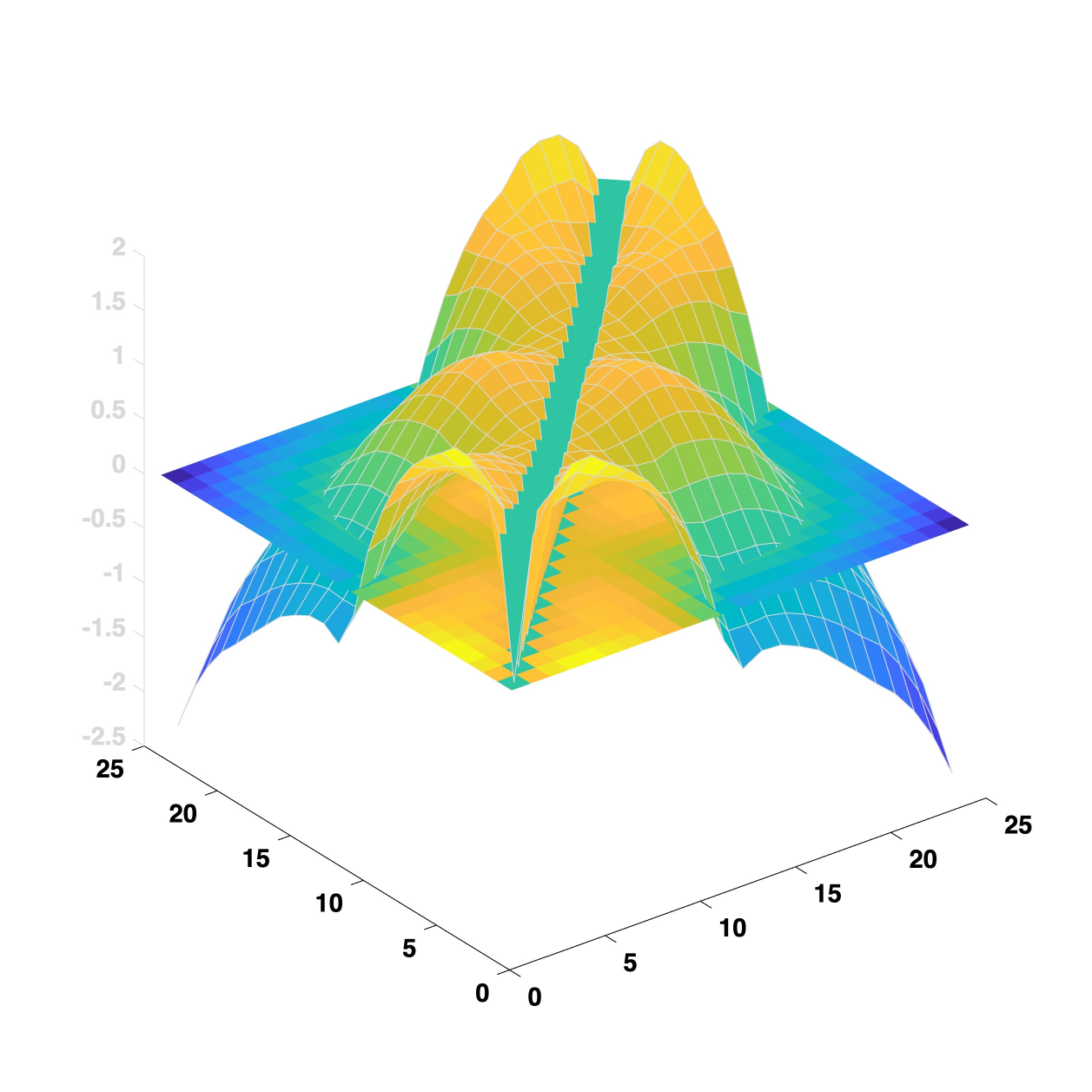}
         \caption{Loop Flow - Group 1}
         \label{fig:sc1}
     \end{subfigure}
     \centering
     \begin{subfigure}[t]{0.45\linewidth}
         \centering
         \includegraphics[width=\textwidth]{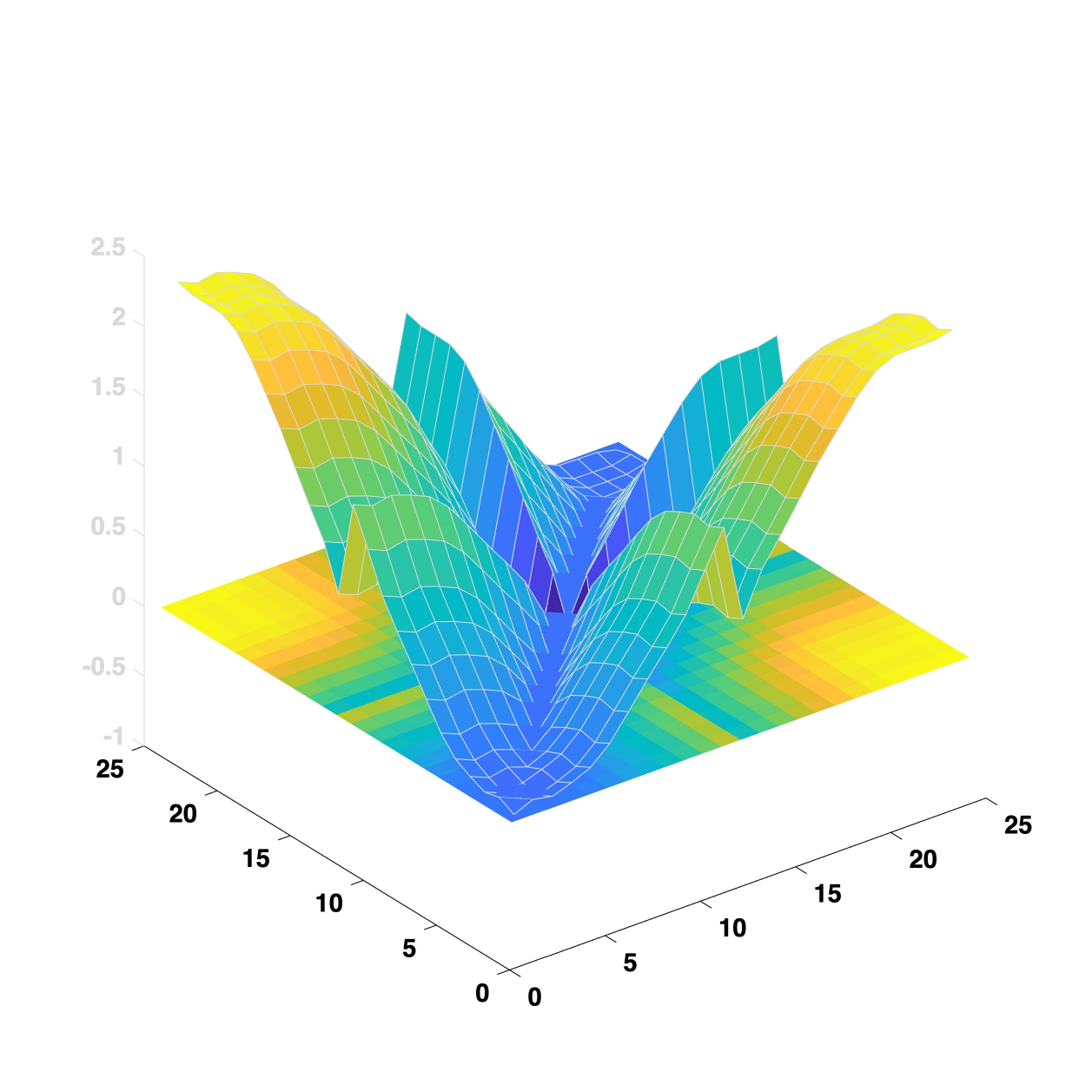}
         \caption{Non-Loop Flow - Group 2}
         \label{fig:sg2}
     \end{subfigure}
     \hfill
     \begin{subfigure}[t]{0.45\linewidth}
         \centering
         \includegraphics[width=\textwidth]{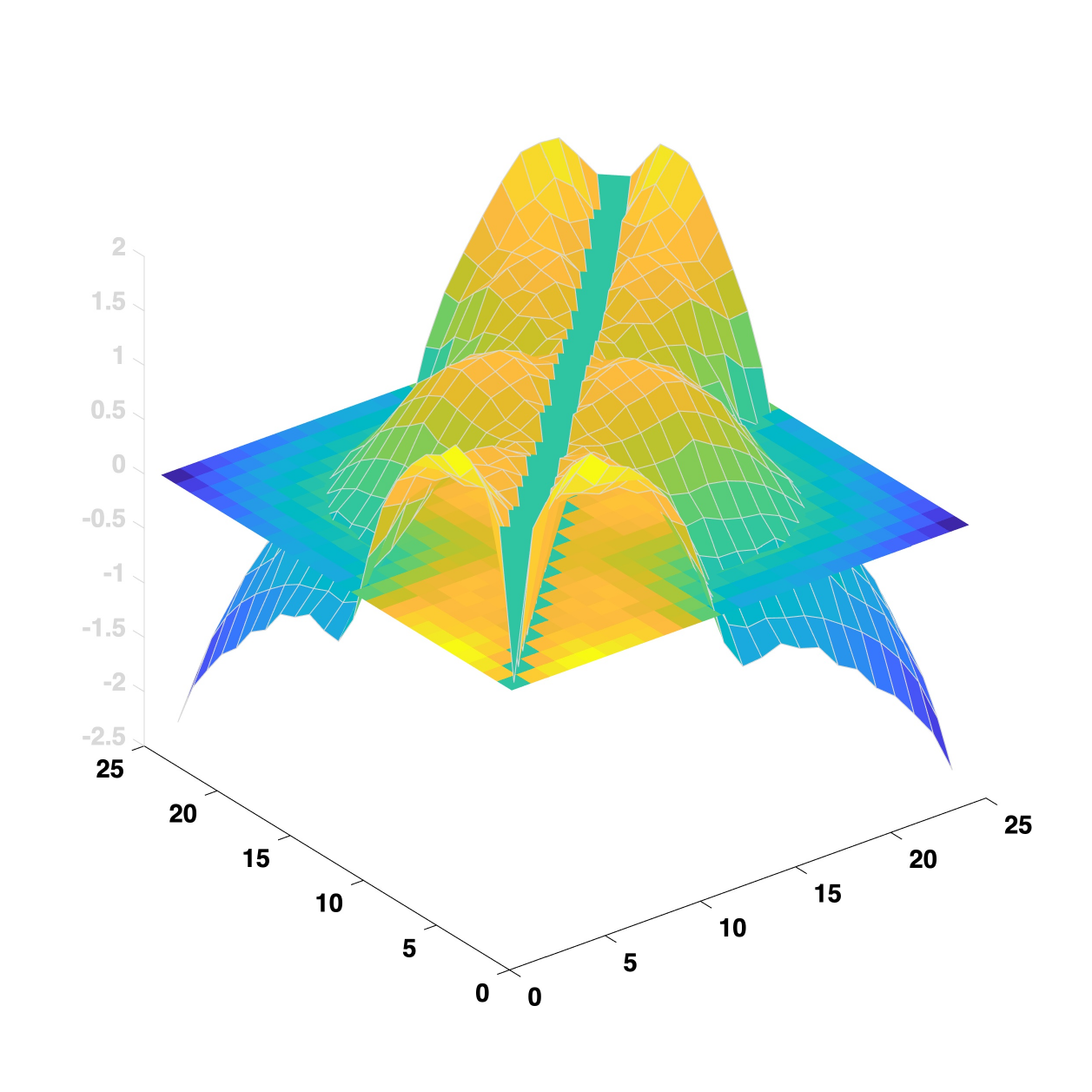}
         \caption{Loop Flow - Group 2}
         \label{fig:sc2}
     \end{subfigure}
\caption{The decomposition of the two groups of networks to be compared. Top: (a), (b) the loop and non-loop flow of Group 1 (Base + Modular). Bottom: (c), (d) the loop and non-loop flow of Group 2 (Base + Modular). The color intensity is an indicator of connectivity strength.}
\label{fig:sim-hdc}
\end{figure}

For each $A^{(j)}_i$, the Hodge decomposition described in Section~\ref{sec:hodge_decompose} is applied to the Base + Modular network. Since the networks generated are complete graphs, there is no harmonic component. We will term the gradient component the loop flow, and the curl component the non-loop flow as a reference to the structure inherent in each decomposition. Figure~\ref{fig:sim-hdc} shows the loop flow and the non-loop flow for the two groups. 

\subsection{Experimental Results}

To quantify separation between the two groups we compute the observed test statistic $\mathcal{T}_{obs}$.
A permutation test with $N=1000$ recomputes the barycenters and statistic on each permuted split of the pooled $2n$ samples, producing $\{T_k\}_{k=1}^{N}$. The Monte Carlo p-value is computed as
\begin{equation*}
\hat p \;=\; \frac{1+\sum_{k=1}^{N}\mathbf{1}\{T_k \ge T\}}{1+N}.
\end{equation*}
The procedure is repeated $50$ independent times, and the average p-value and standard deviation is reported in Table~\ref{tab:sim-comp}. Different number of networks from 5 to 50 were generated for each group. Larger values are preferred across all group comparison as that will indicate that the variance minimization is effective at removing small perturbations.
\begin{table*}[t!]
	\setlength{\tabcolsep}{3.5pt}
	\centering
    \renewcommand{\arraystretch}{1.5}
	\begin{tabular}{c|c|ccc}
		\centering
		Decomp. & Groups & 5 networks & 25 networks & 50 networks\\
		\hline
	  \multirow{3}{*}{Edge Flow} 
        & 1 vs. 1 & 0.2775 (0.1762) & 0.2637 (0.1809) & 0.2495 (0.1736) \\ 
		& 2 vs. 2 & 0.3159 (0.1879) & 0.2565 (0.1554) & 0.2555 (0.1644) \\ 
        & 1 vs. 2 & 0.2313 (0.1795) & \cellcolor{Gray} 0.0261 (0.0685) & \cellcolor{Gray}0.0007 (0.0037) \\ 
        \hline
	 \multirow{3}{*}{Non-Loop Flow} 
        & 1 vs. 1 & 0.2456 (0.1793) & 0.2281 (0.1470) & 0.2388 (0.1674) \\ 
		& 2 vs. 2 & 0.3128 (0.1918) & 0.2900 (0.1560) & 0.2565 (0.2009) \\ 
        & 1 vs. 2 & \cellcolor{Gray} 0.2382 (0.2007) & \cellcolor{Gray} 0.1214 (0.1484) &  0.0228 (0.0606) \\ 
        \hline
	 \multirow{3}{*}{Loop Flow} 
        & 1 vs. 1 & 0.2585 (0.1859) & 0.2591 (0.1646) & 0.2470 (0.1500) \\ 
		& 2 vs. 2 & 0.2746 (0.2024) & 0.2564 (0.1538) & 0.2862 (0.1586) \\ 
        & 1 vs. 2 & \cellcolor{Gray} 0.2107 (0.1898) & 0.0397 (0.0539) & 0.0340 (0.0972) \\ 
		\hline
	\end{tabular}	
    \caption{The performance results on differentiating networks of different topologies using the Wasserstein variance minimization technique. Different number of networks from 5 to 50 were generated for each group. Larger p-values are preferred across all comparisons since the goal is to show that the Wasserstein variance minimization method can accurately filter the effect of noise on networks.}
	\label{tab:sim-comp}
\end{table*}

For networks within the same group, the test produced consistently high p-values across all three Hodge components, supporting topological equivalence within homogeneous sets. When comparing networks from different groups, the Loop Flow and Non-Loop Flow components generally yield larger p-values for the smaller sample size, but their p-values decline as the number of networks increases, indicating the variance minimization often reduces between group differences at small sample sizes but becomes less pronounced with larger samples. By contrast, the Edge Flow shows a large p-value only in the 5 networks comparison between groups, while for 25 and 50 networks the p-values are small, indicating statistically detectable differences. Because the Edge Flow corresponds to the original, unprojected network structure, it is more sensitive to small perturbations and therefore less robust for group comparisons. These results indicate that applying Wasserstein variance minimization to the loop and non loop components of the Hodge decomposition can suppress noise driven variance and promote comparability across groups, whereas the raw edge component is susceptible to small perturbations and can reveal group differences when sample size increases even when that difference difference is due to small perturbations.

These simulation results validate two key properties of the proposed framework. First, the Hodge decomposition isolates network structure into components with different sensitivities to perturbations: Non-Loop Flows are more robust to local noise, while Loop Flows and especially raw edges are more susceptible. Second, Wasserstein variance minimization in the Non-Loop Flow subspace provides the strongest noise filtering, maintaining high p-values even at moderate sample sizes where edge-based methods yield false positives. This makes the decomposed approach particularly valuable for comparing networks where measurement noise or biological variability creates small random fluctuations that should not be interpreted as meaningful topological differences.

\section{Application to Functional Brain Networks}
\label{sec:app}

The framework is applied to a subset of the Addiction Connectome Preprocessed Initiative (ACPI) distribution of the \emph{Multimodal Treatment of ADHD (MTA)} resting-state fMRI resource released through FCP/INDI. The next section describes the minimal preprocessing pipeline for the data.

\subsection{Data}
The dataset used for this study comes from the Addiction Connectome Preprocessed Initiative (ACPI), which provides a preprocessed version of the \emph{Multimodal Treatment of ADHD (MTA)} resting-state fMRI dataset distributed via FCP/INDI. This release includes curated raw scans that have been standardized into 4D volumes and summarized as region-level time series. For this analysis, we utilized the parcellated time series based on the Automated Anatomical Labeling (AAL) atlas \cite{tzourio2002automated}. Participants were divided into two groups according to marijuana use status: individuals who regularly used cannabis (users) and those who did not (non-users).  

The ACPI framework implements a factorial preprocessing scheme that varies along three specific dimensions following standard motion correction and normalization to MNI space. These factors include the spatial normalization algorithm (\textit{ANTS} or \textit{FNIRT}), whether high-motion frames were censored, and whether global signal regression was applied. Consequently, each subject may have up to eight possible preprocessing variants, all parcellated within the same AAL atlas. To maintain consistency, a single variant was selected \emph{a priori} for all participants to prevent mixing preprocessing conditions across groups. For each subject, the ROI time series were mean-centered, and Pearson correlations were computed between all ROI pairs to construct symmetric AAL connectivity matrices. Correlation coefficients were Fisher $z$–transformed prior to computing group averages and then back-transformed for reporting and visualization. No additional preprocessing steps such as denoising, filtering, spatial smoothing, or regression were applied beyond those in the ACPI pipeline. This approach ensures transparency in preprocessing choices while enabling consistent parcel-level functional network representations for subsequent edge-based analyses.
\begin{figure*}[ht]
     \centering
     \begin{subfigure}[t]{0.325\linewidth}
         \centering
         \includegraphics[width=\textwidth]{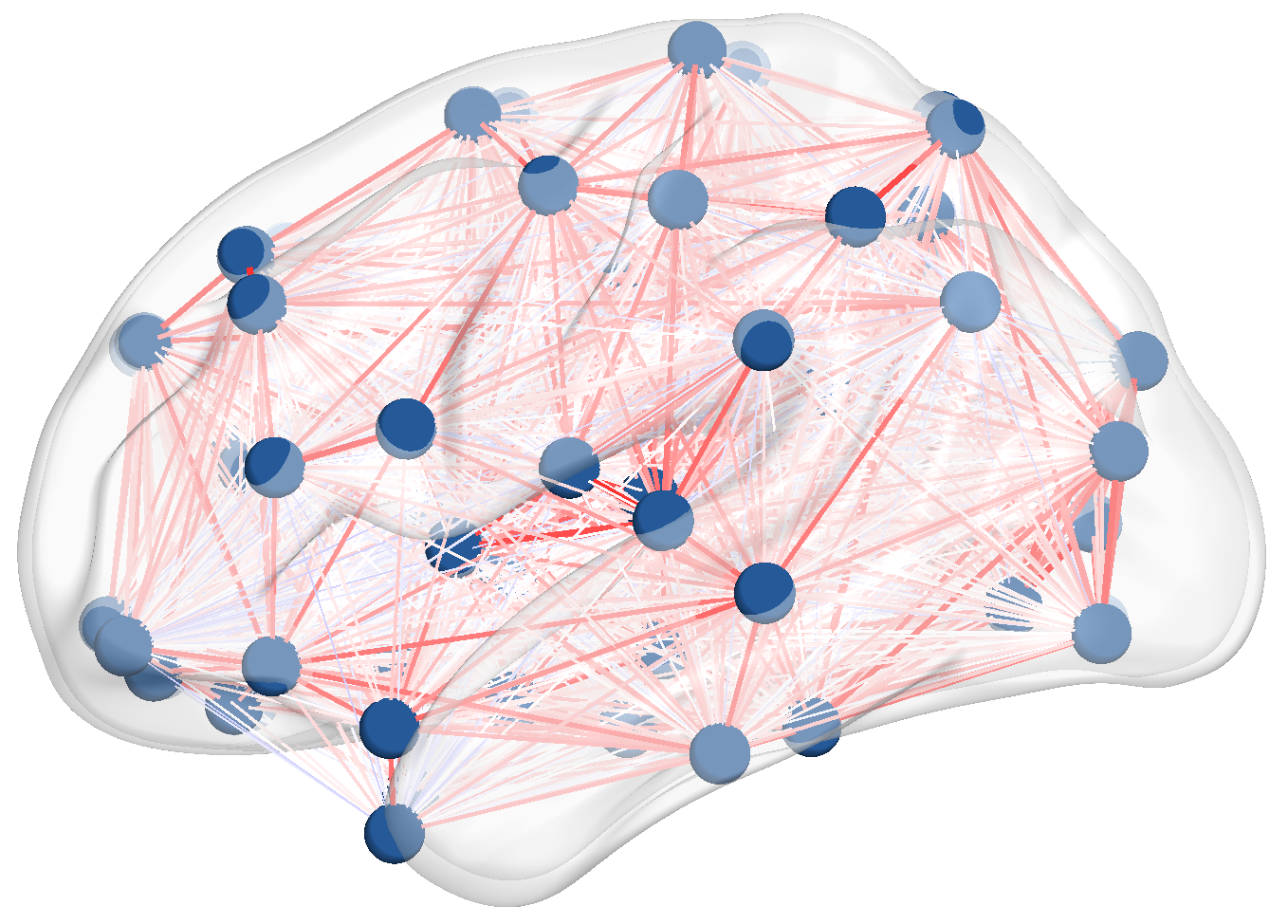}
     \end{subfigure}
     \hfill
     \begin{subfigure}[t]{0.325\linewidth}
         \centering
         \includegraphics[width=0.75\textwidth,height=0.75\textwidth]{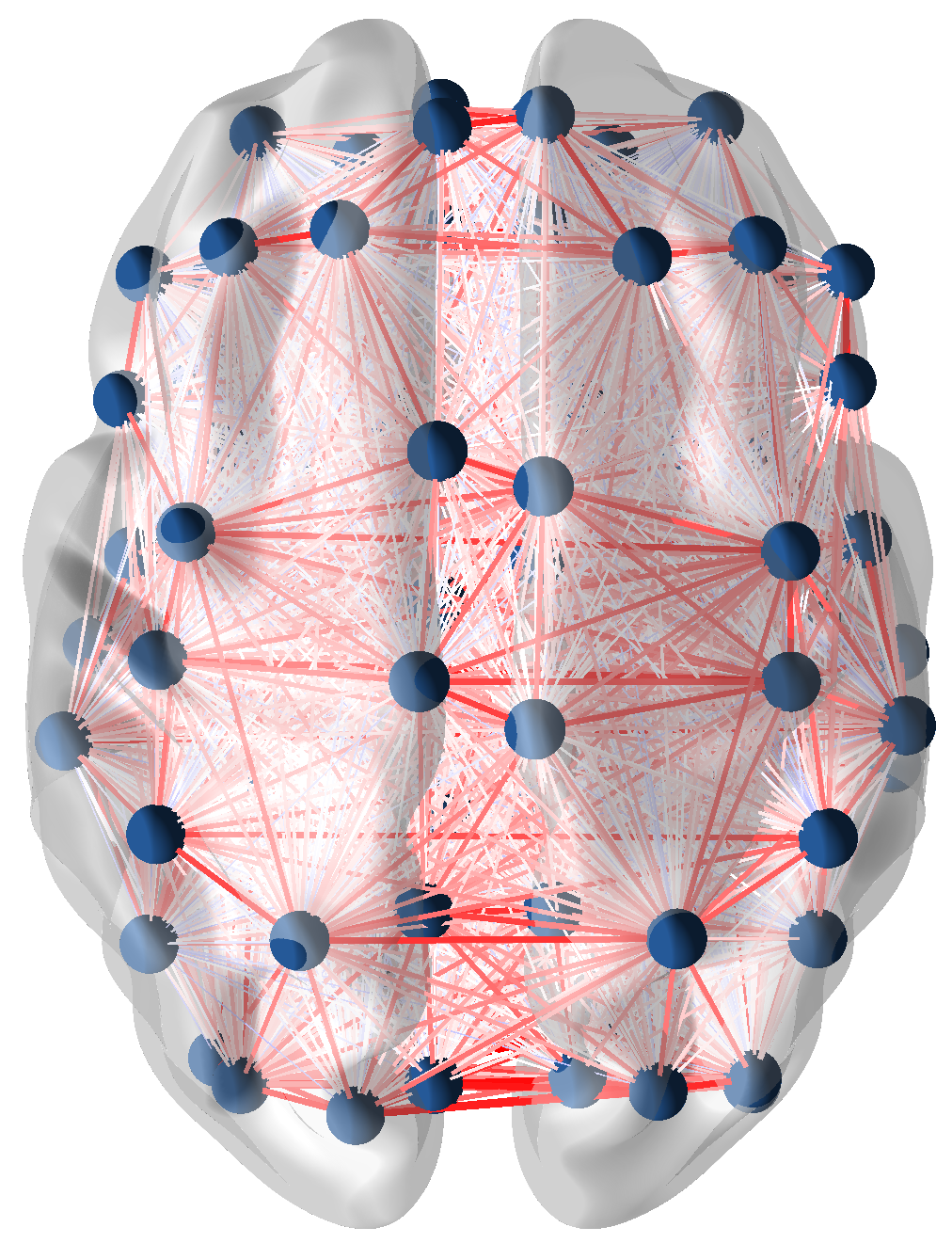}
     \end{subfigure}
     \hfill
     \begin{subfigure}[t]{0.325\linewidth}
         \centering
         \includegraphics[width=\textwidth]{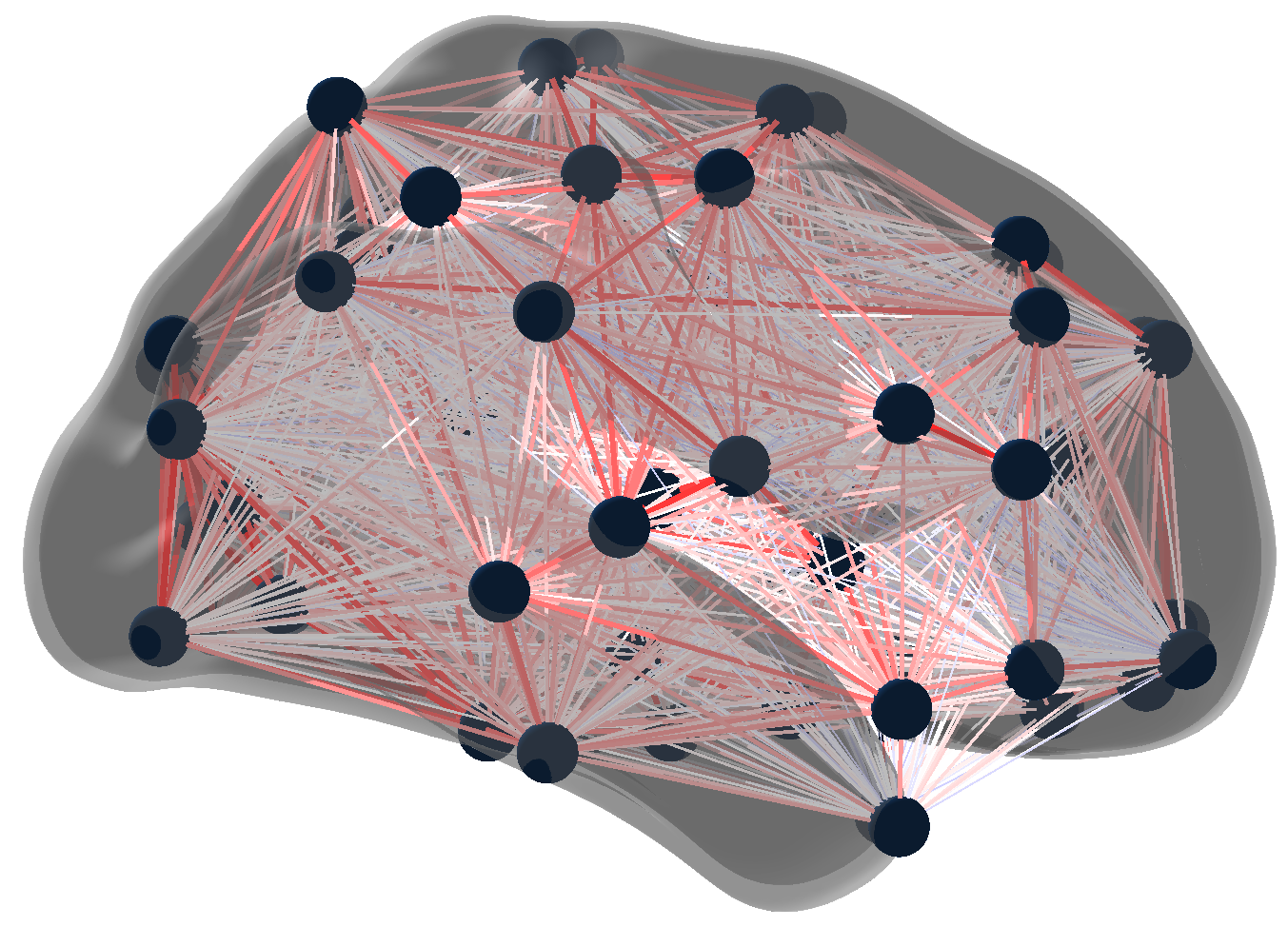}
     \end{subfigure}
     
    \par\vspace{0.25cm}
    
     \centering
     \begin{subfigure}[t]{0.325\linewidth}
         \centering
         \includegraphics[width=\textwidth]{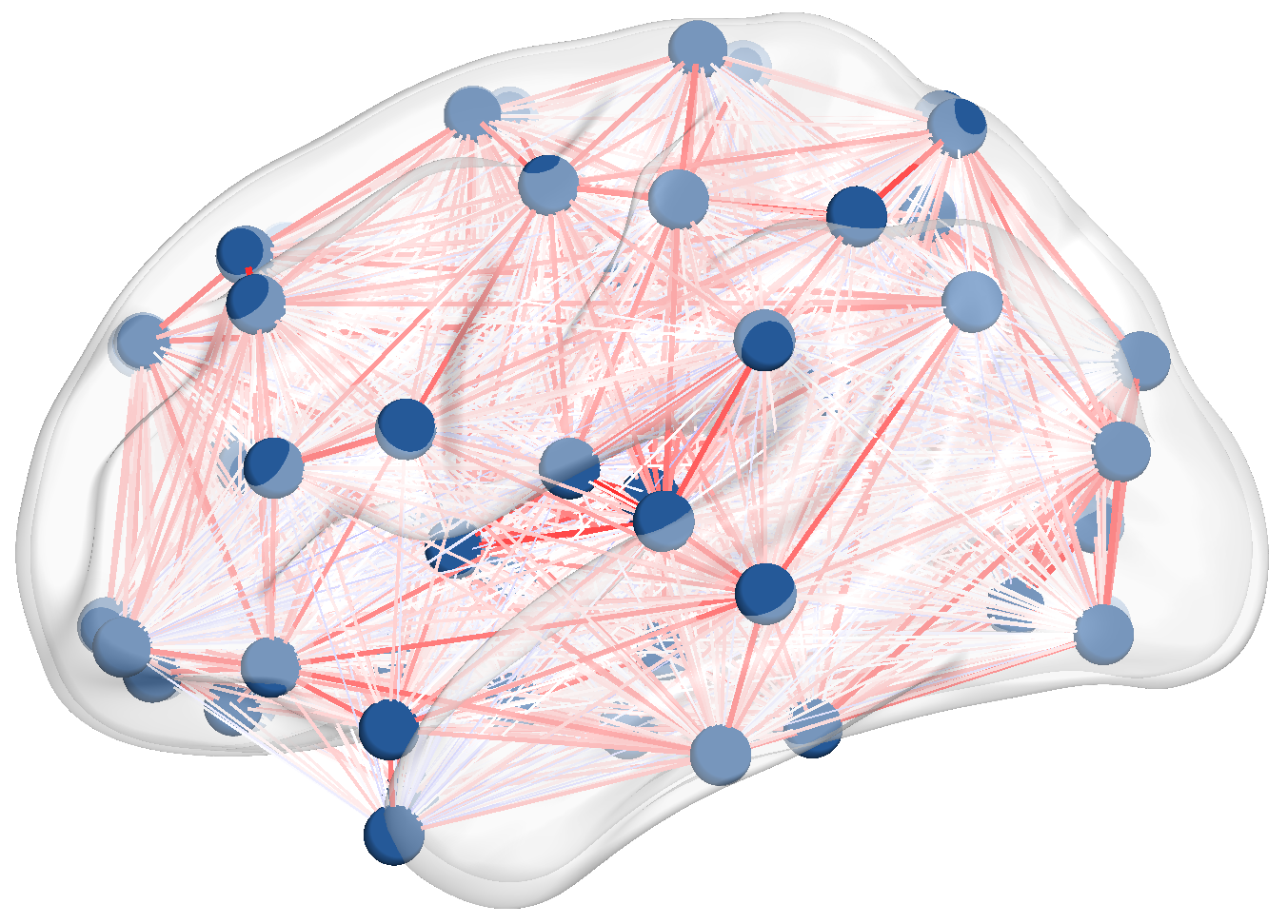}
     \end{subfigure}
     \hfill
     \begin{subfigure}[t]{0.325\linewidth}
         \centering
         \includegraphics[width=0.75\textwidth,height=0.75\textwidth]{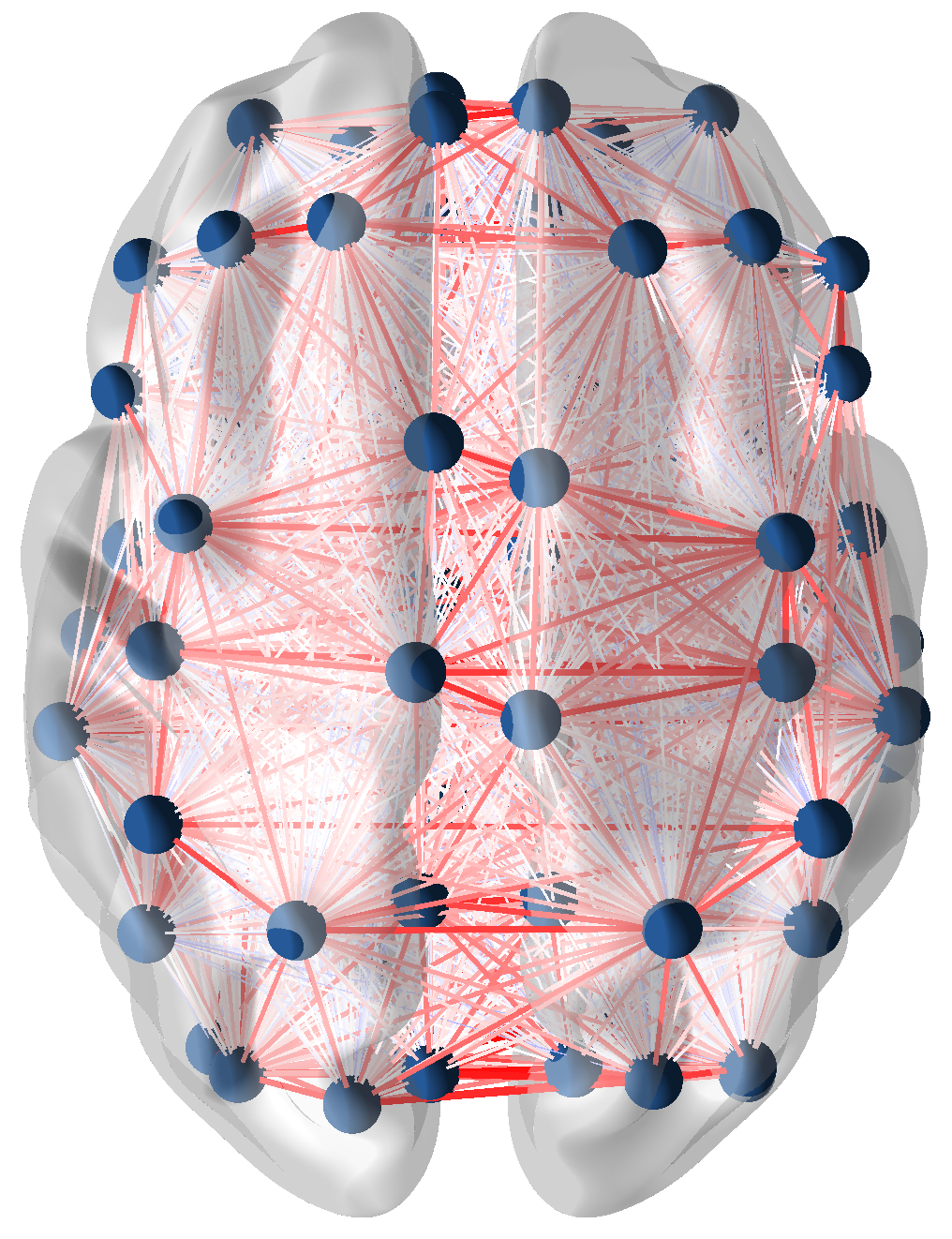}
     \end{subfigure}
     \hfill
     \begin{subfigure}[t]{0.325\linewidth}
         \centering
         \includegraphics[width=\textwidth]{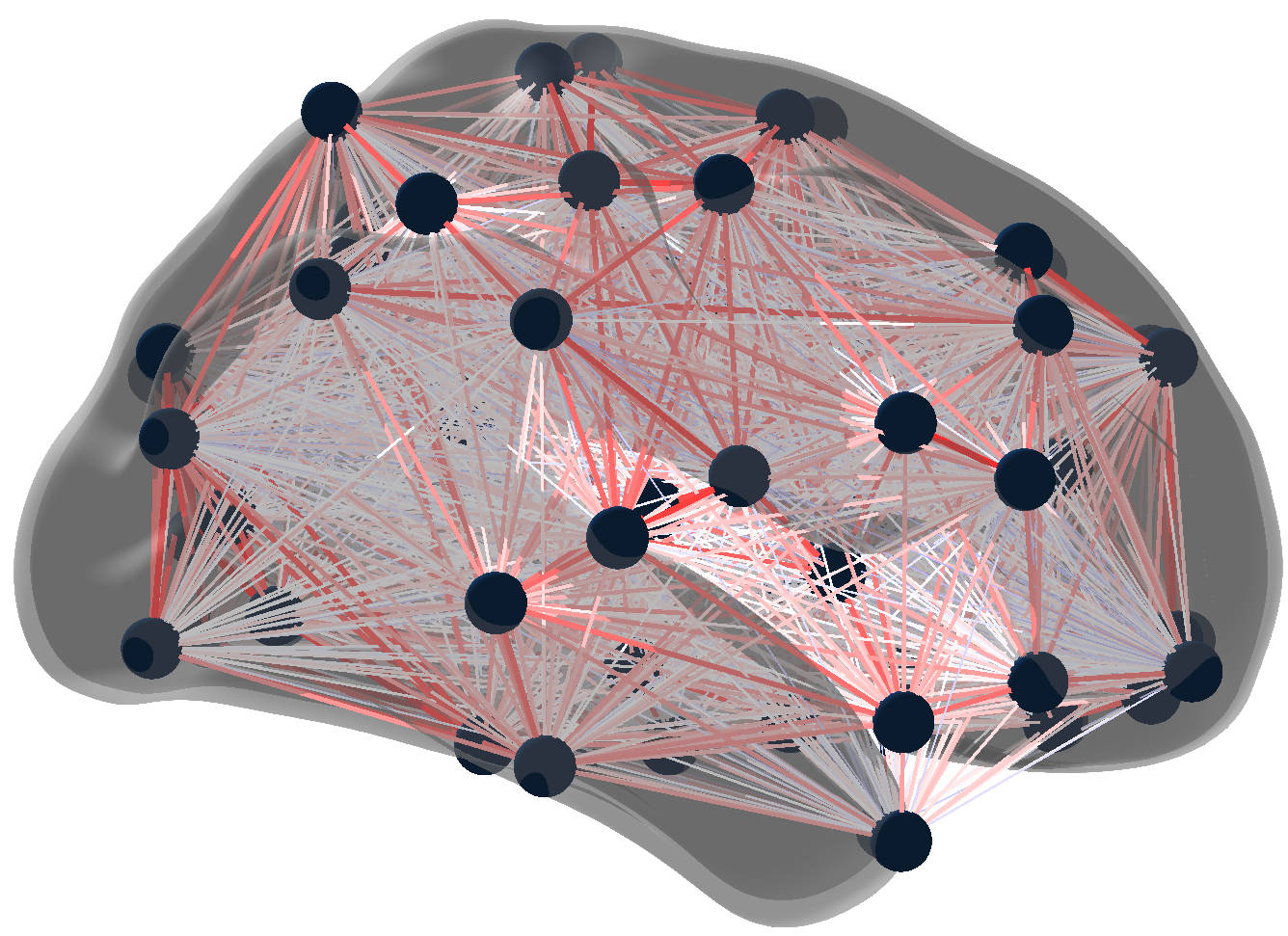}
     \end{subfigure}
      \begin{subfigure}[t]{1\linewidth}
         \centering
         \includegraphics[width=0.75\textwidth]{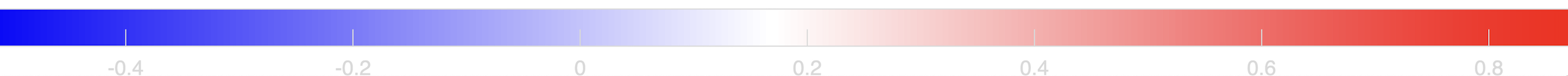}
     \end{subfigure}
    \caption{Top: The Edge flow of non-users. Relatively few negative connections were observed. Bottom: The Edge Flow of users. Relatively few negative connections were observed.}
    \label{fig:edge-flow}
\end{figure*}
The spatial distributions of the original networks (Edge Flow) are illustrated in Figures~\ref{fig:edge-flow}, showing left hemisphere, dorsal, and right hemisphere views for both user and non-user groups across all three flow components.

\subsection{Results}

\begin{figure}[ht!]
     \centering
     \begin{subfigure}[t]{0.325\linewidth}
         \centering
         \includegraphics[width=\textwidth]{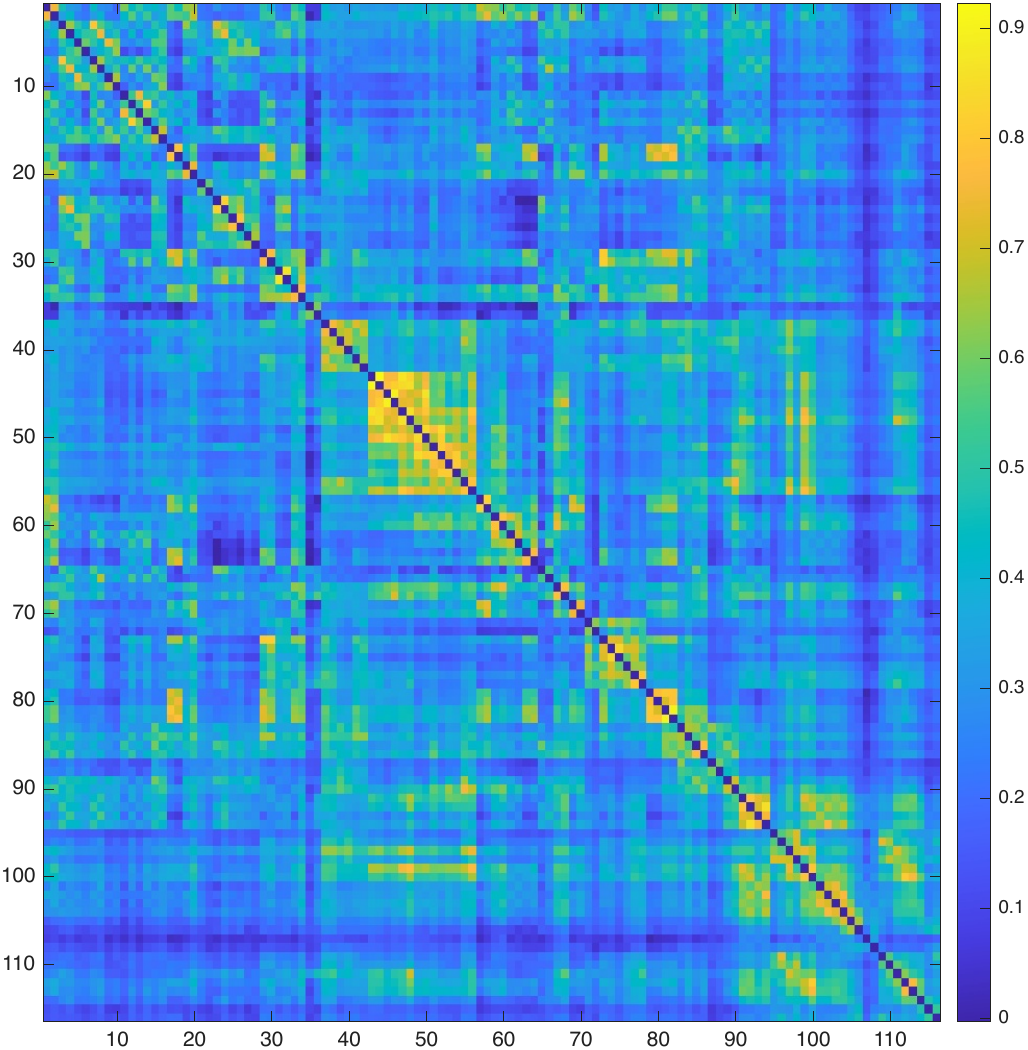}
         \caption{Edge Flow: Non-users}
         \label{fig:e1}
     \end{subfigure}
     \hfill
     \begin{subfigure}[t]{0.325\linewidth}
         \centering
         \includegraphics[width=\textwidth]{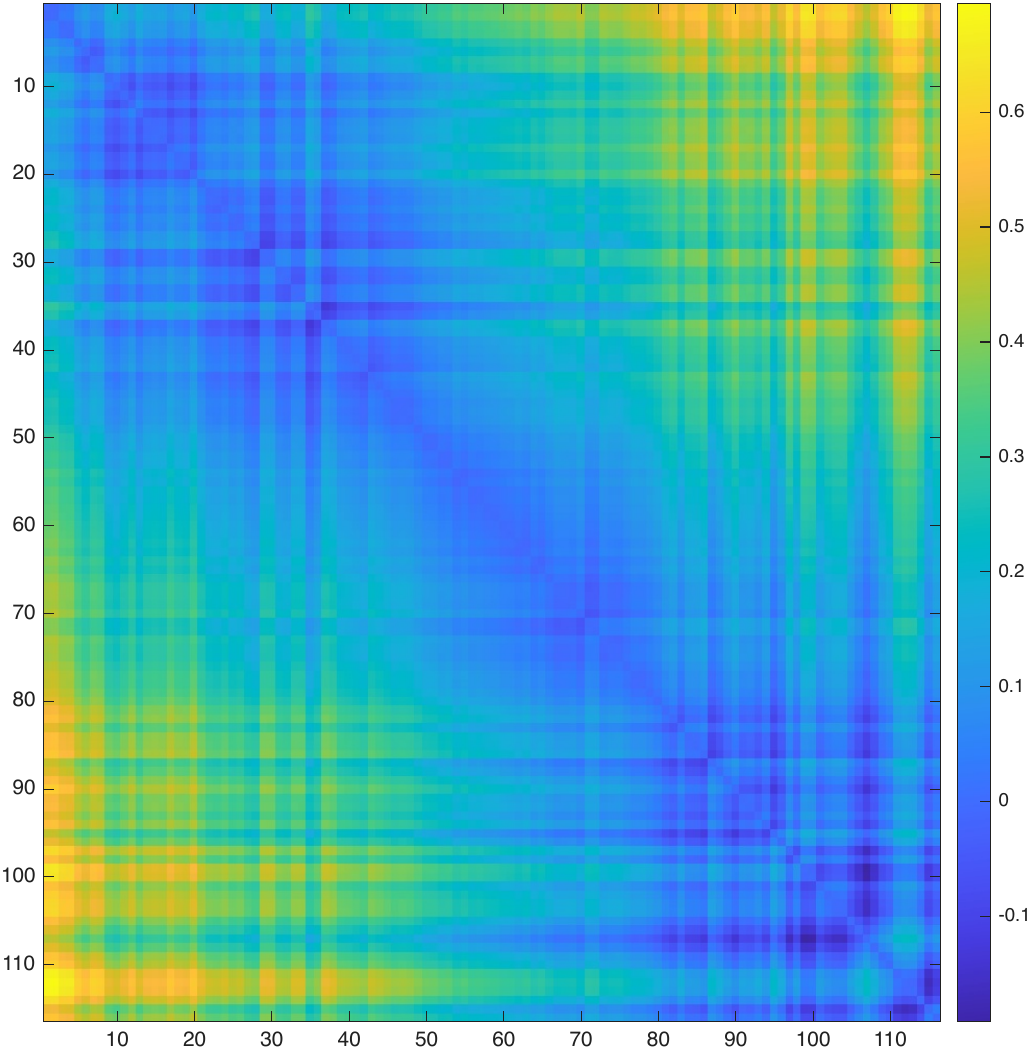}
         \caption{Non-Loop Flow: Non-users}
         \label{fig:g1}
     \end{subfigure}
     \hfill
     \begin{subfigure}[t]{0.325\linewidth}
         \centering
         \includegraphics[width=\textwidth]{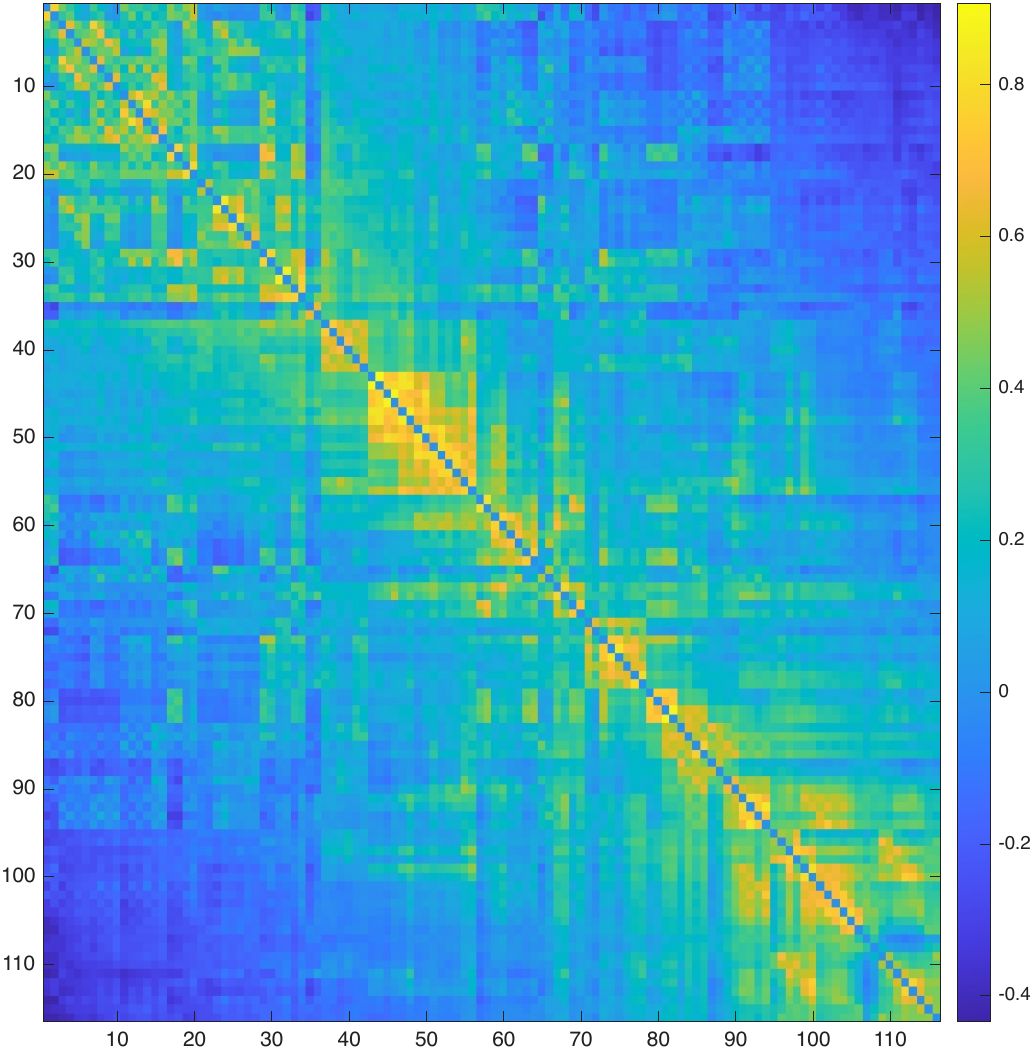}
         \caption{Loop Flow: Non-Users}
         \label{fig:c1}
     \end{subfigure}
     \centering
     \begin{subfigure}[t]{0.325\linewidth}
         \centering
         \includegraphics[width=\textwidth]{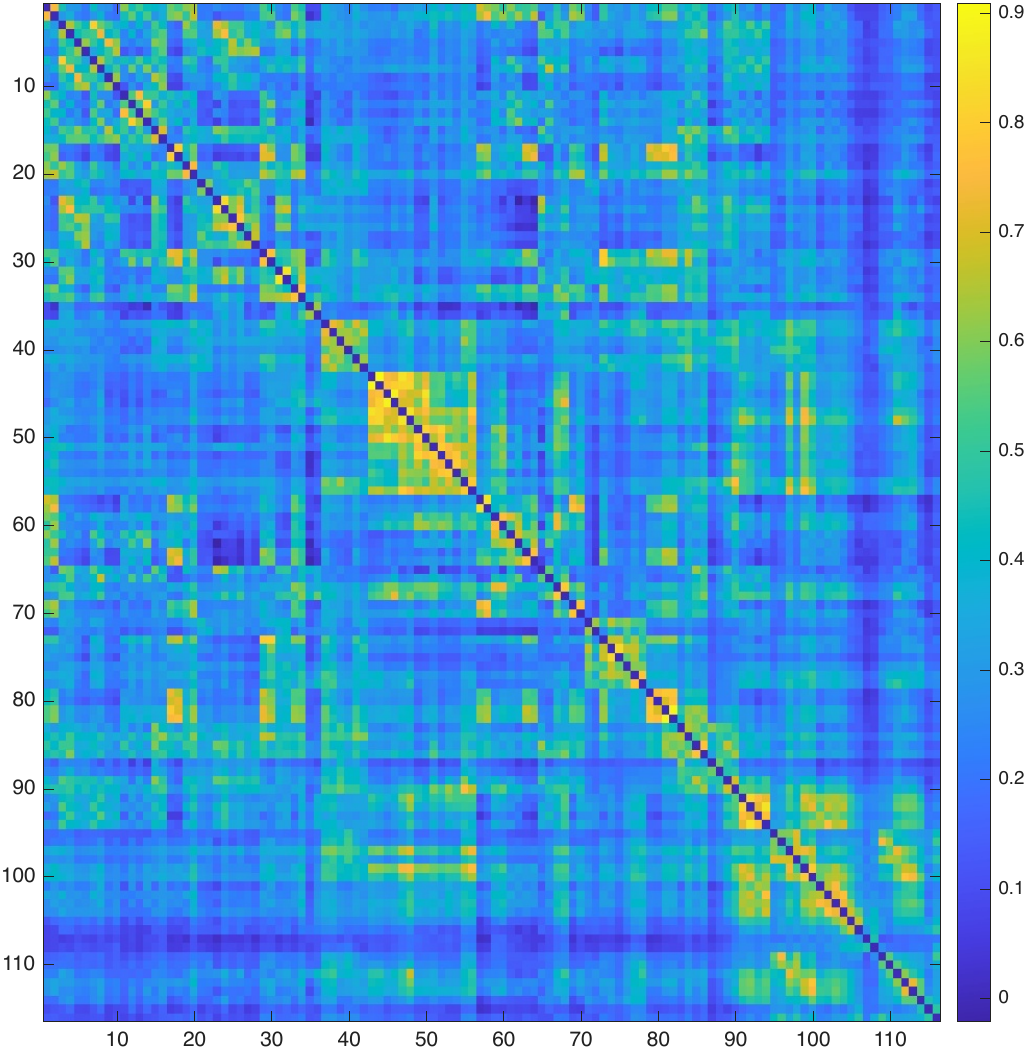}
         \caption{Edge Flow: Users}
         \label{fig:e2}
     \end{subfigure}
     \hfill
     \begin{subfigure}[t]{0.325\linewidth}
         \centering
         \includegraphics[width=\textwidth]{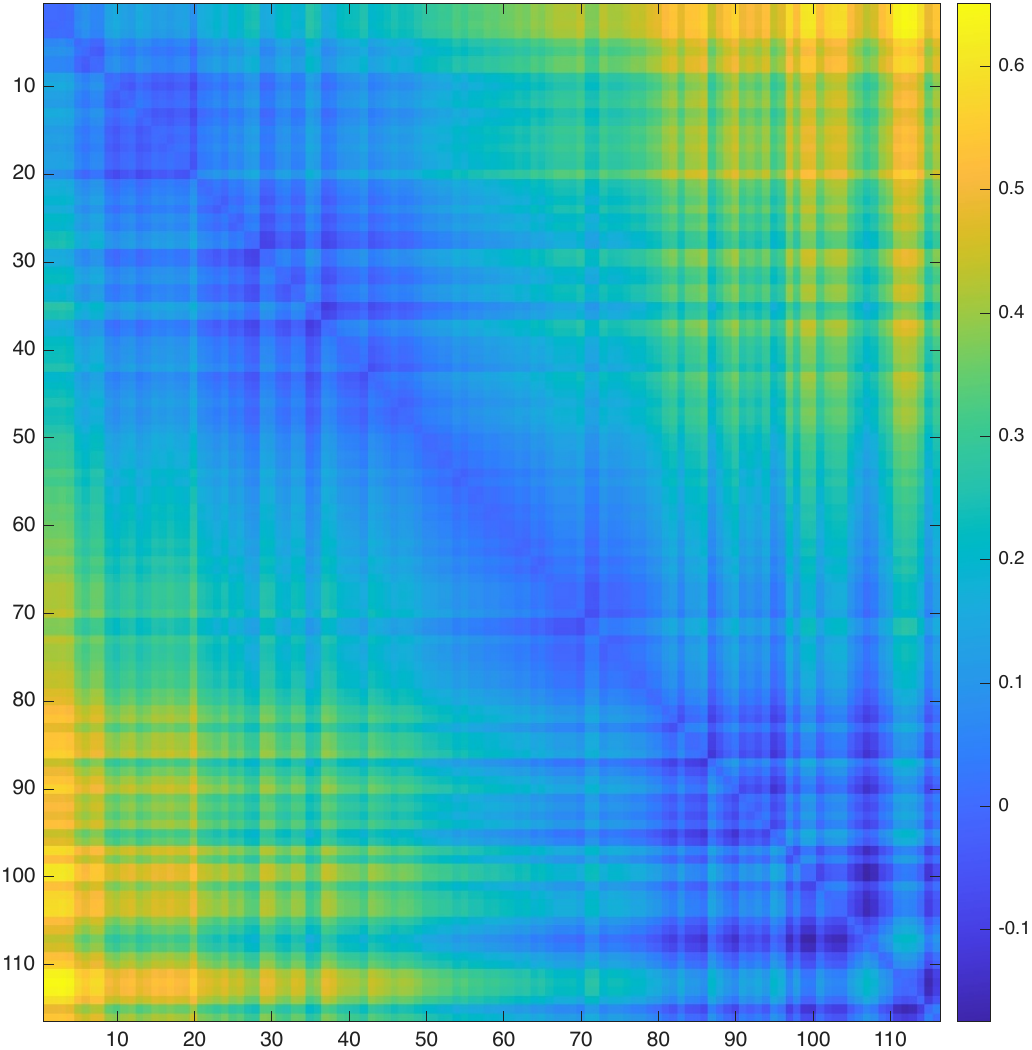}
         \caption{Non-Loop Flow: Users}
         \label{fig:g2}
     \end{subfigure}
     \hfill
     \begin{subfigure}[t]{0.325\linewidth}
         \centering
         \includegraphics[width=\textwidth]{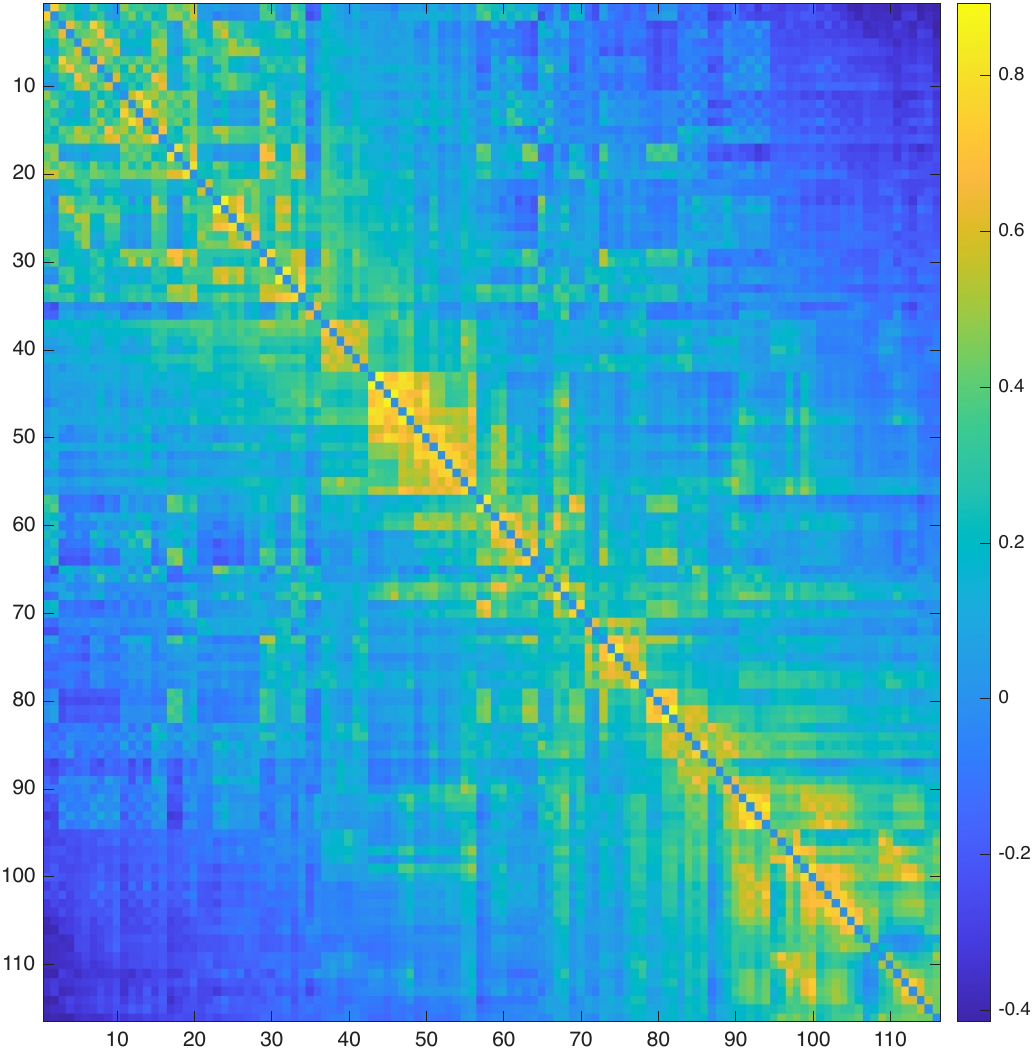}
         \caption{Loop: Users}
         \label{fig:c2}
     \end{subfigure}
\caption{Hodge decomposition of the Edge Flow averaged across all subjects. Left: The Edge Flow for non-users and users. Middle: The Non-Loop Flow for users and non-users. Right: The Loop Flow for users and non-users.}
    \label{fig:app-dcmp}
\end{figure}

The group-averaged connectivity matrices and their Hodge decomposition into loop and non-loop flows are shown in Figure~\ref{fig:app-dcmp}. The aggregate edge flow (Figures~\ref{fig:e1} and~\ref{fig:e2}) shows widespread positive correlations throughout cortex in both groups, with relatively few negative connections. The connectivity matrices reveals subtle differences in the spatial patterning of connections, particularly in the non-loop flow where users exhibit altered distributions of local connectivity compared to non-users.

\begin{figure*}[ht]
     \centering
     \begin{subfigure}[t]{0.325\linewidth}
         \centering
         \includegraphics[width=\textwidth]{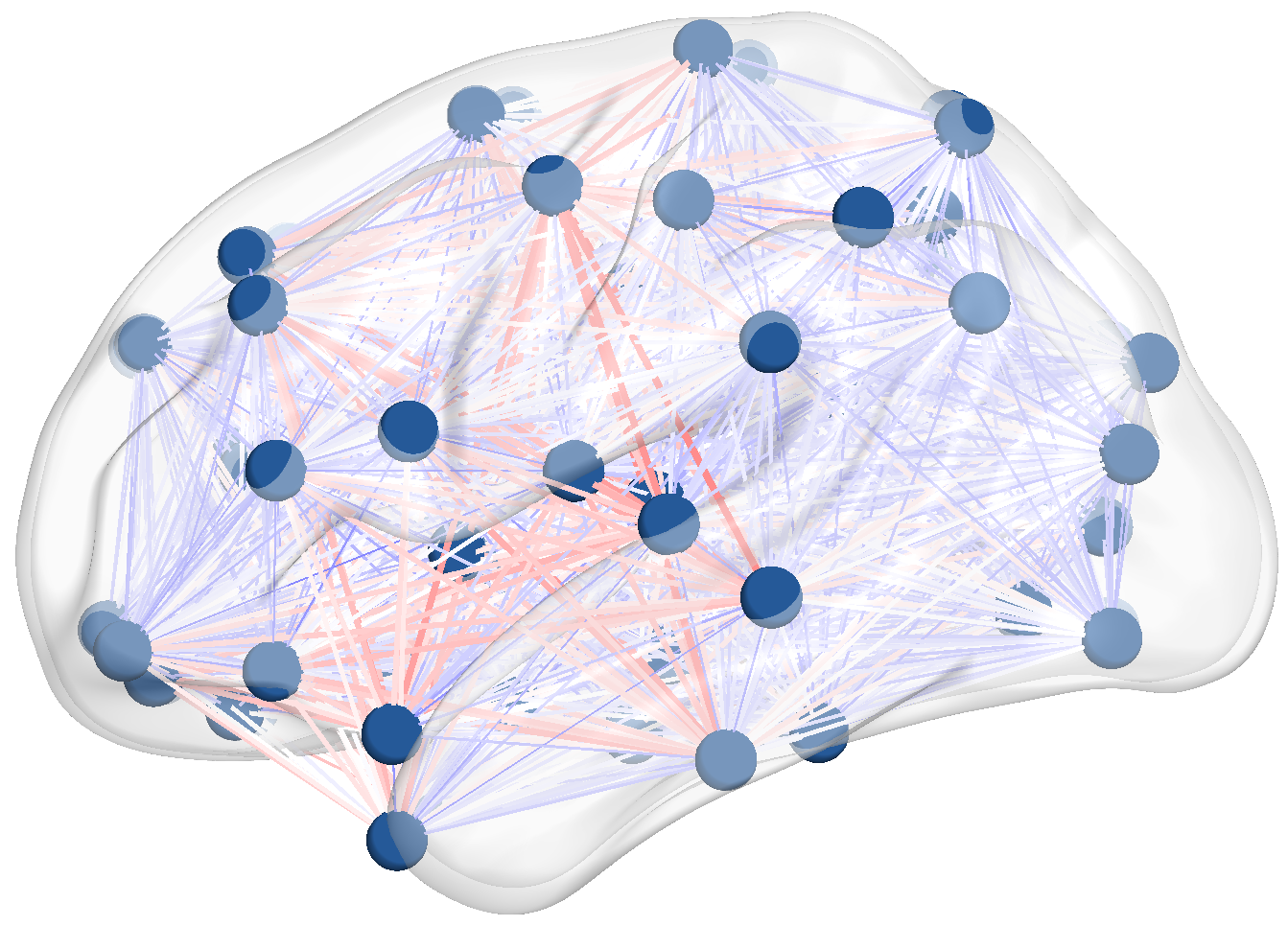}
     \end{subfigure}
     \hfill
     \begin{subfigure}[t]{0.325\linewidth}
         \centering
         \includegraphics[width=0.75\textwidth,height=0.75\textwidth]{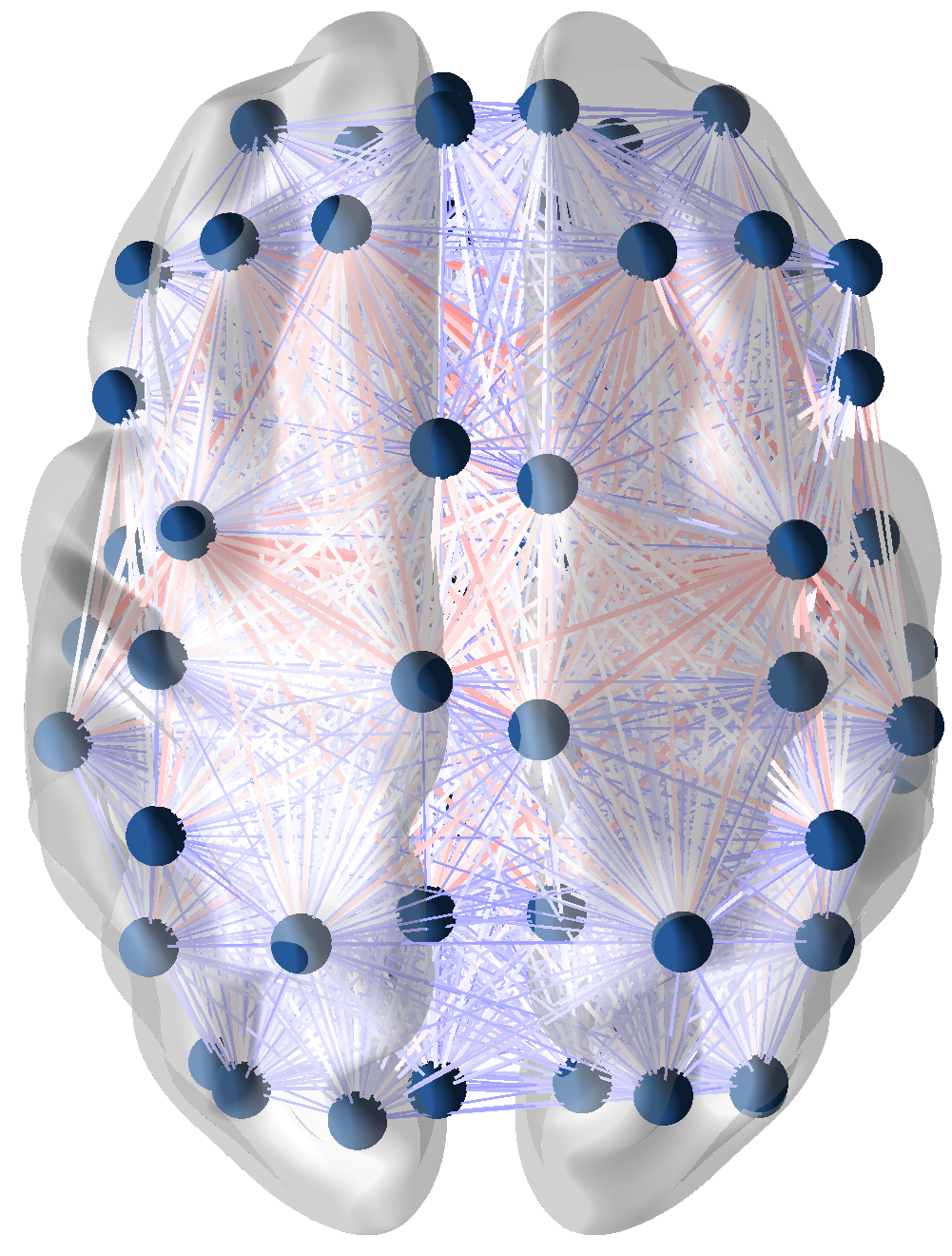}
     \end{subfigure}
     \hfill
     \begin{subfigure}[t]{0.325\linewidth}
         \centering
         \includegraphics[width=\textwidth]{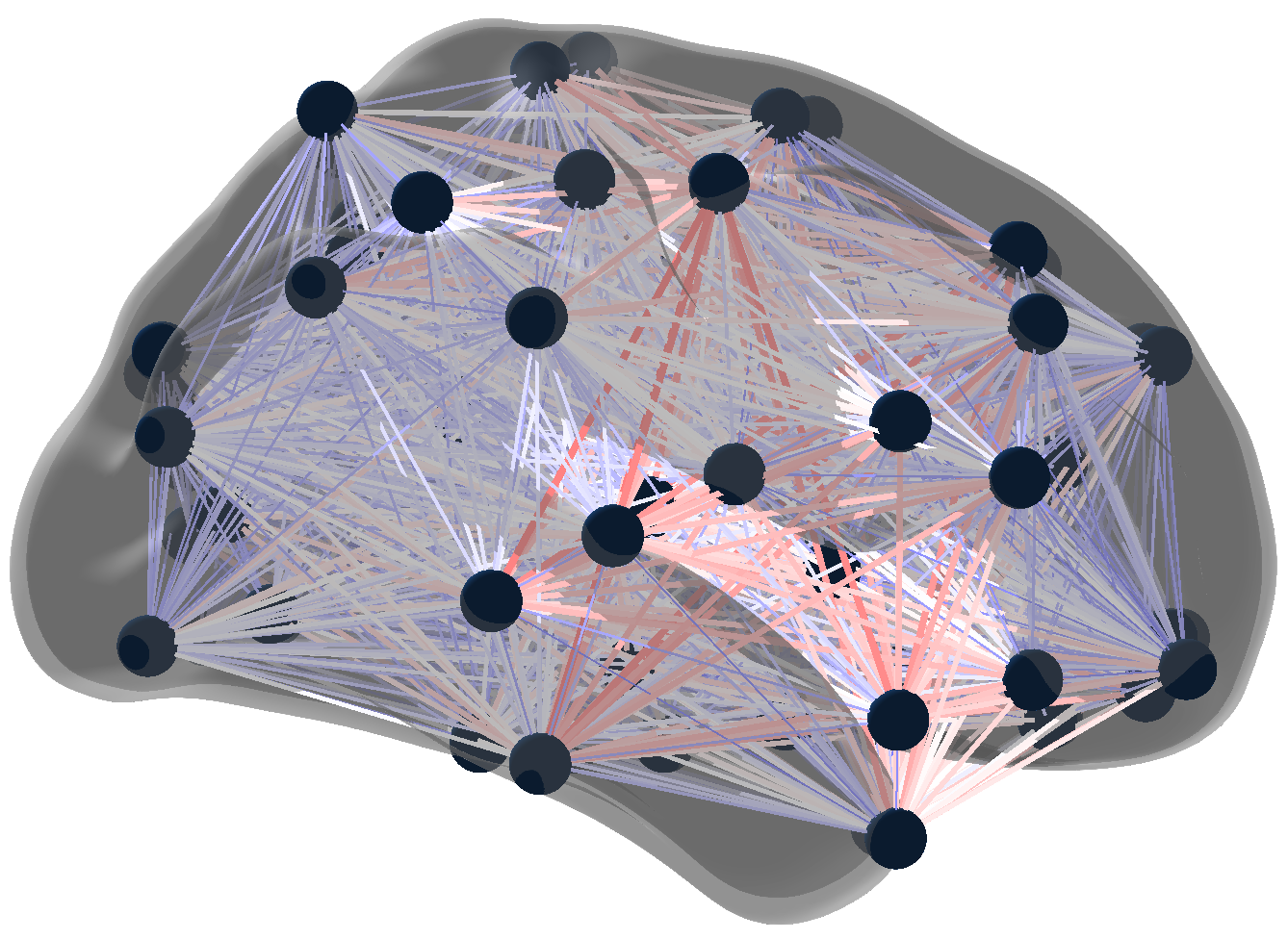}
     \end{subfigure}
     
    \par\vspace{0.25cm}
    
     \centering
     \begin{subfigure}[t]{0.325\linewidth}
         \centering
         \includegraphics[width=\textwidth]{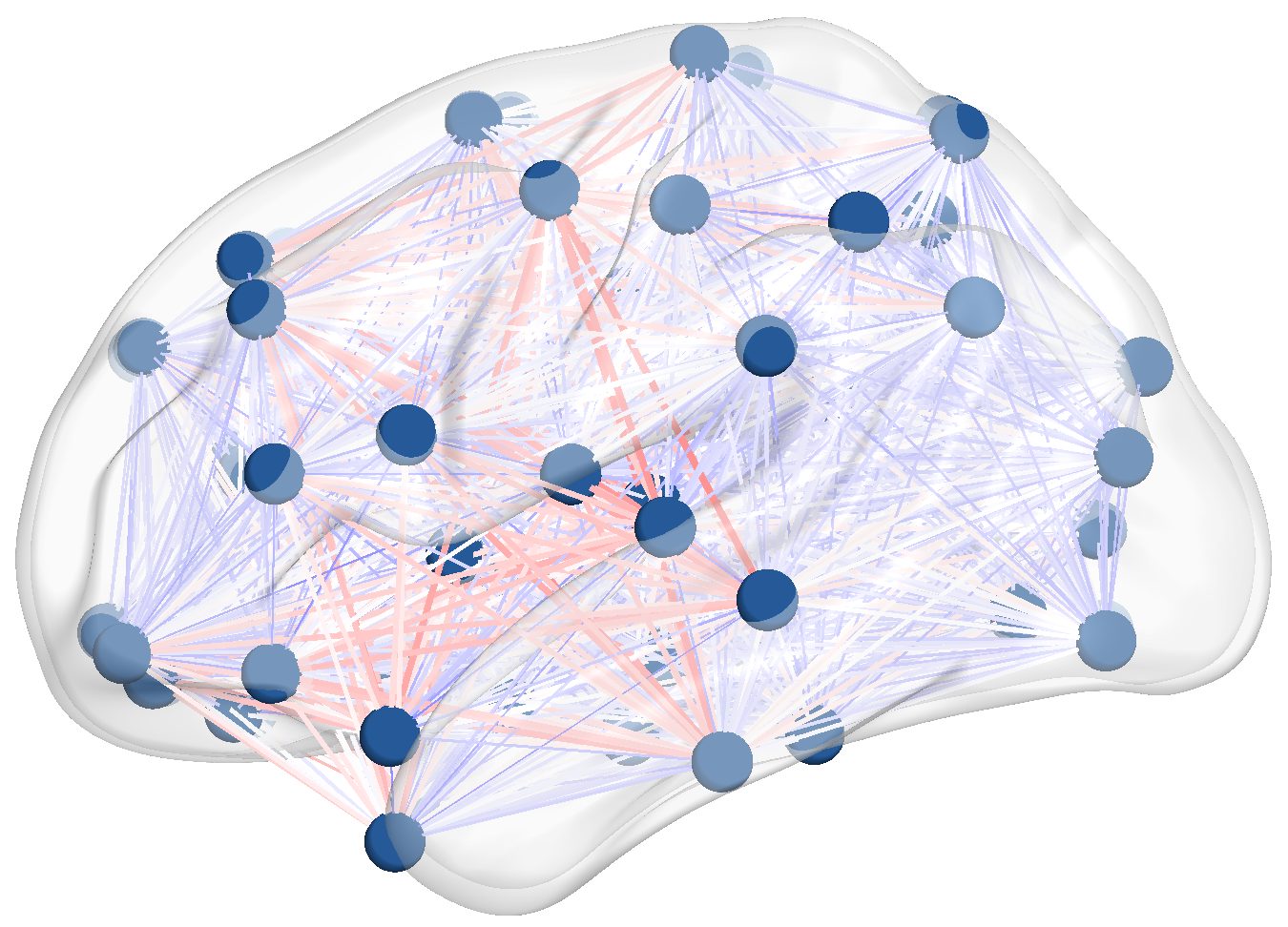}
     \end{subfigure}
     \hfill
     \begin{subfigure}[t]{0.325\linewidth}
         \centering
         \includegraphics[width=0.75\textwidth,height=0.75\textwidth]{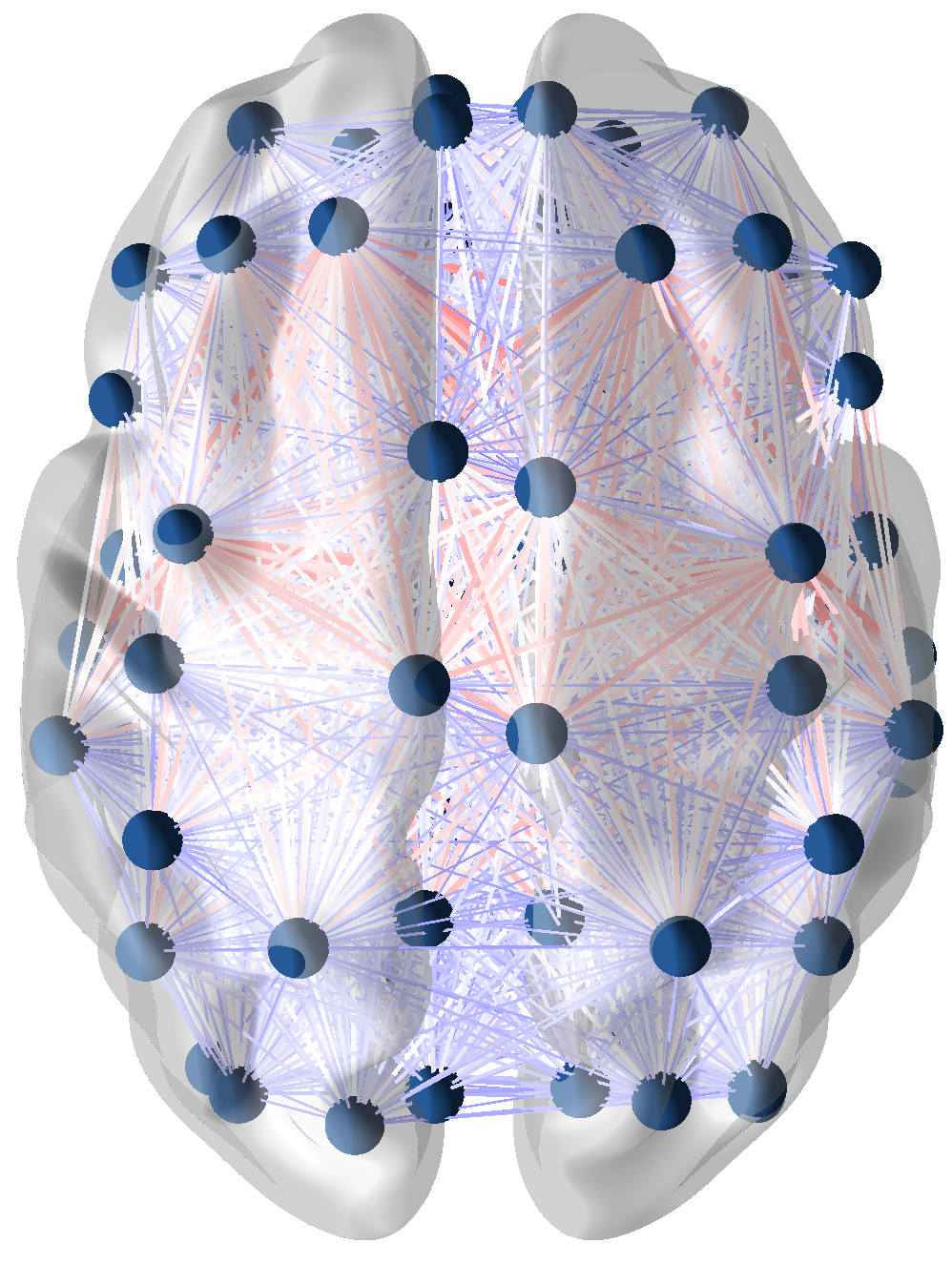}
     \end{subfigure}
     \hfill
     \begin{subfigure}[t]{0.325\linewidth}
         \centering
         \includegraphics[width=\textwidth]{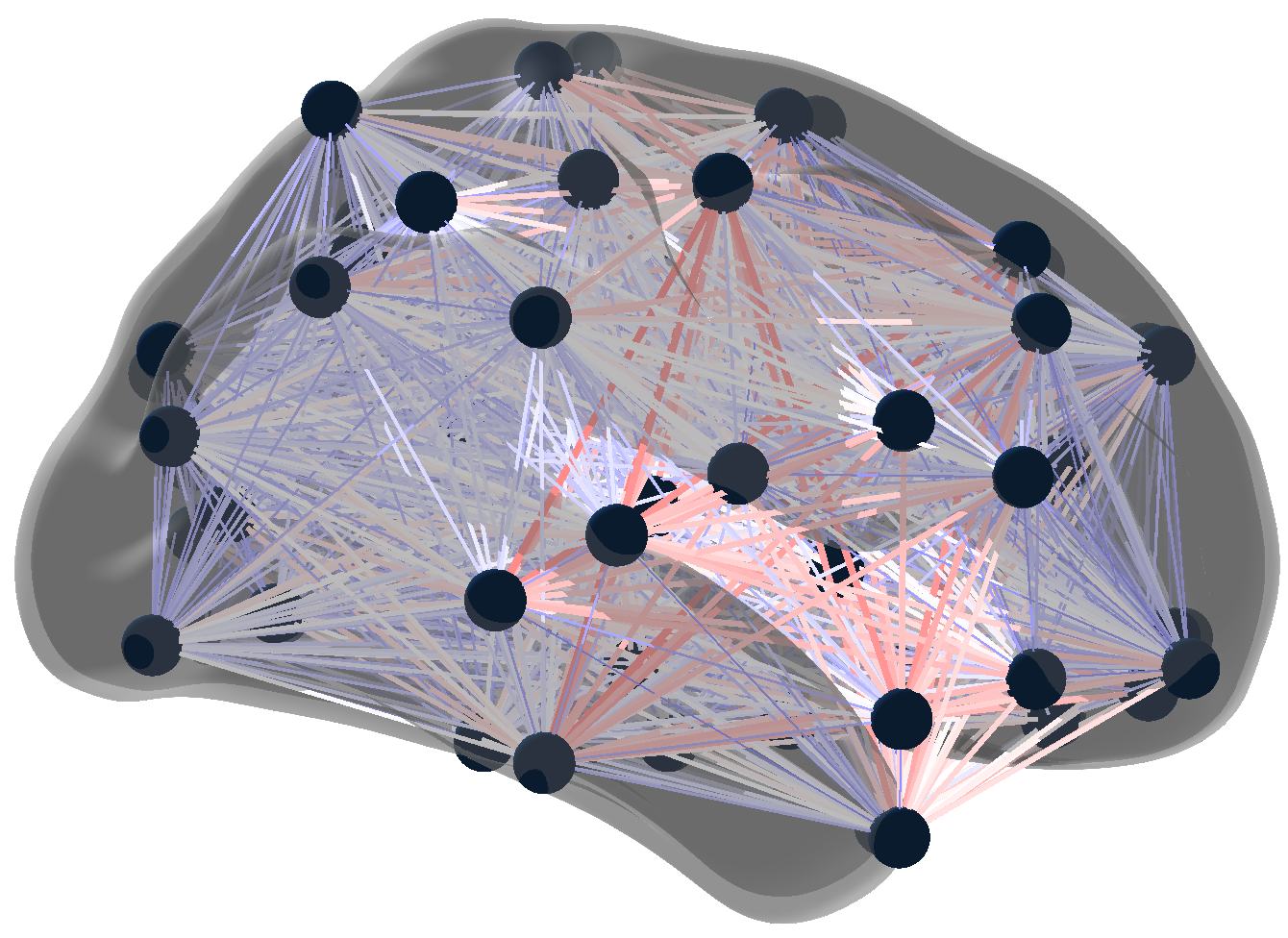}
     \end{subfigure}
      \begin{subfigure}[t]{1\linewidth}
         \centering
         \includegraphics[width=0.75\textwidth]{cb_use1.png}
     \end{subfigure}
    \caption{Top: The Non-Loop Flow of non-users. Relatively few negative connections were observed. Bottom: The Non-Loop Flow of users. Relatively few negative connections were observed. Observe that the networks are more similar in structure across the positive connections. }
    \label{fig:grad-flow}
\end{figure*}

\begin{figure*}[ht]
     \centering
     \begin{subfigure}[t]{0.325\linewidth}
         \centering
         \includegraphics[width=\textwidth]{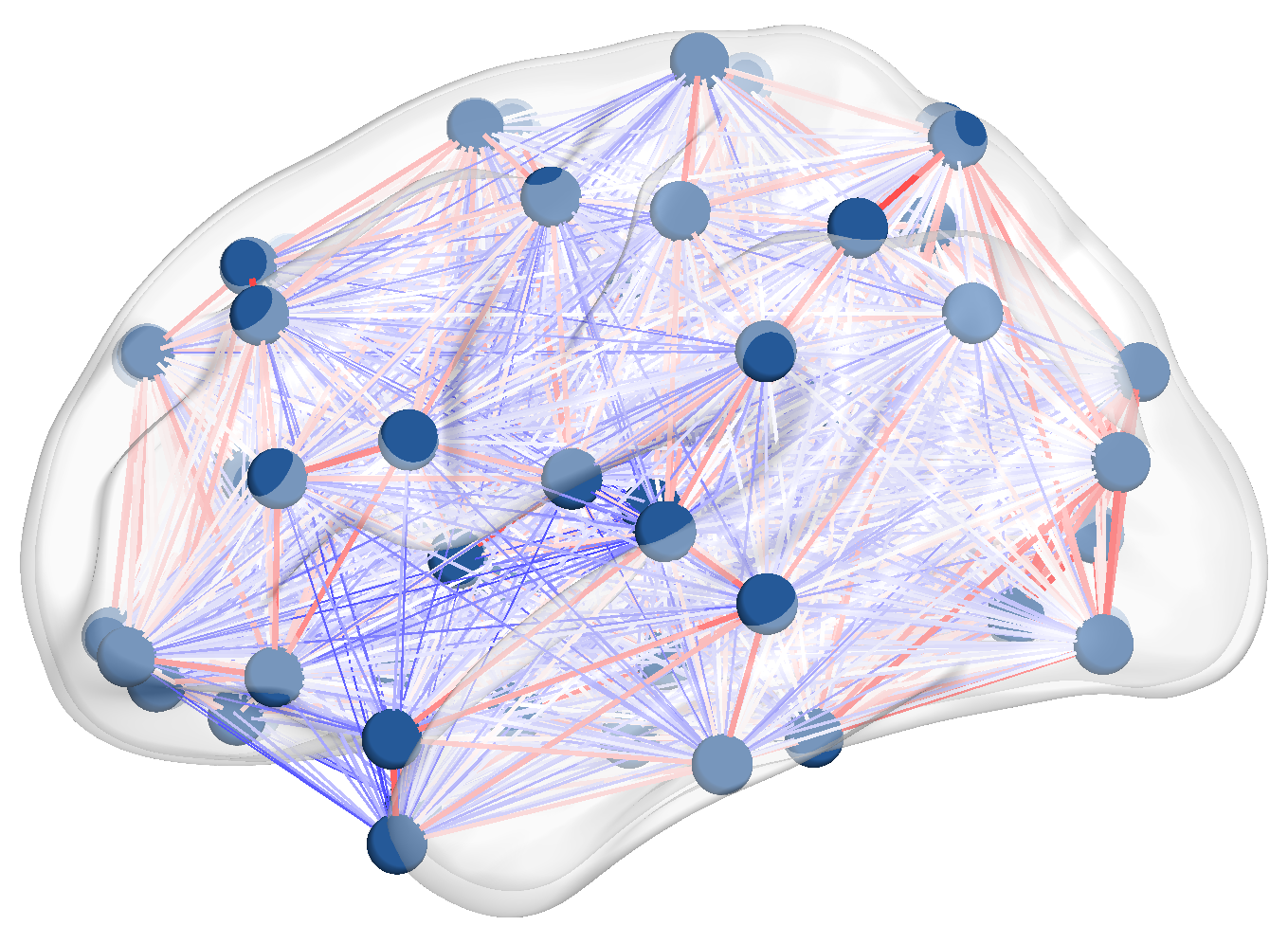}
     \end{subfigure}
     \hfill
     \begin{subfigure}[t]{0.325\linewidth}
         \centering
         \includegraphics[width=0.75\textwidth,height=0.75\textwidth]{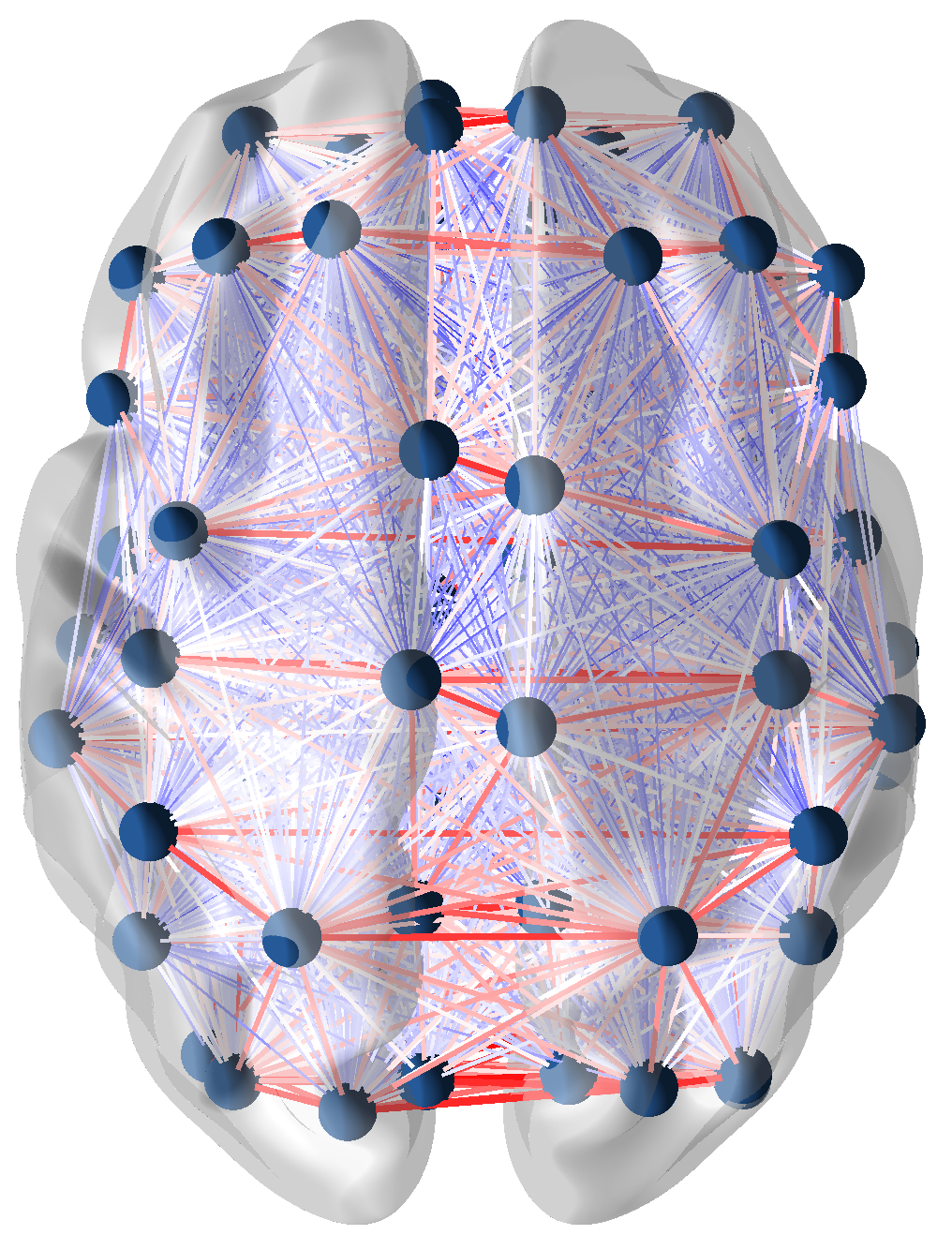}
     \end{subfigure}
     \hfill
     \begin{subfigure}[t]{0.325\linewidth}
         \centering
         \includegraphics[width=\textwidth]{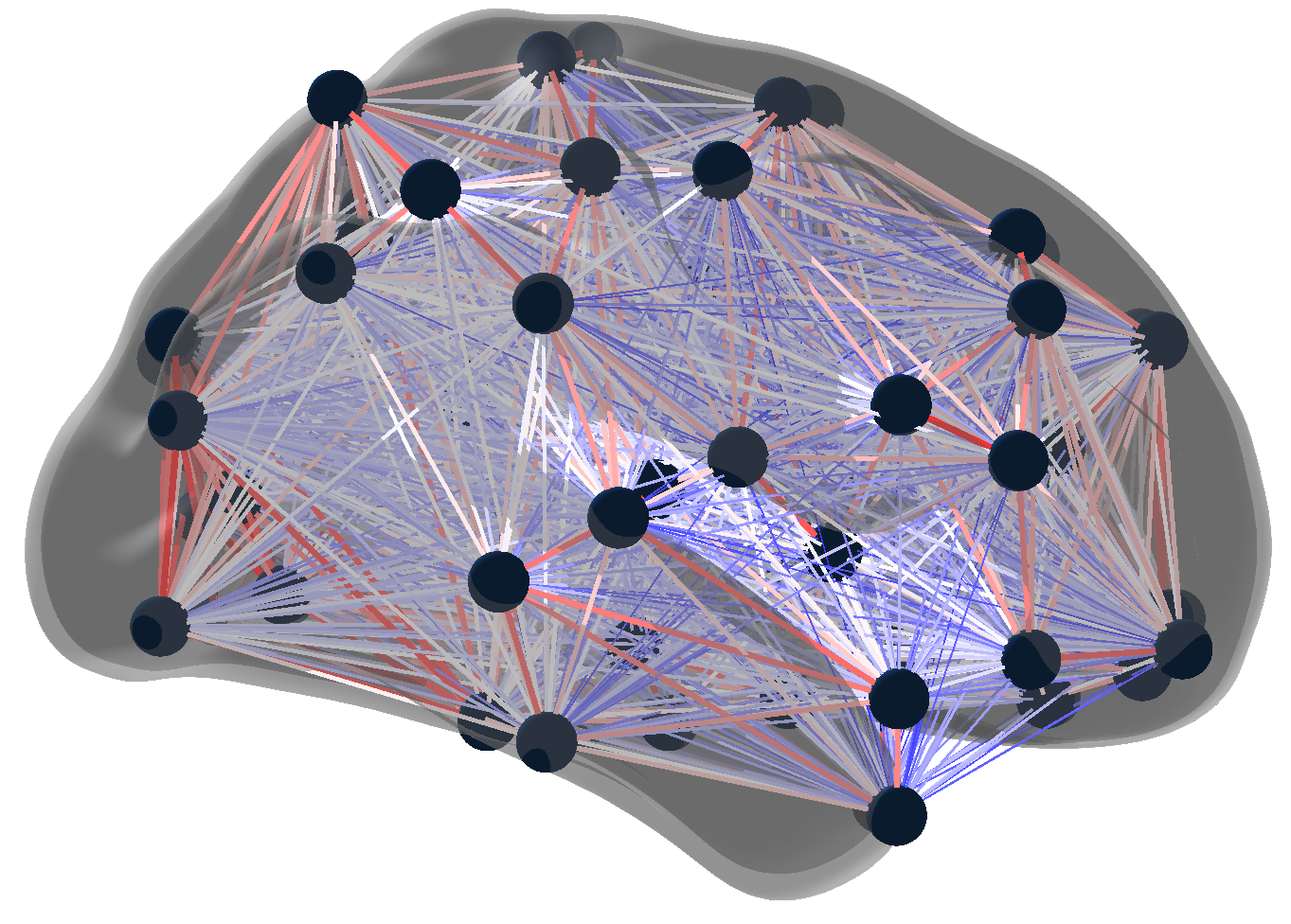}
     \end{subfigure}
     
    \par\vspace{0.25cm}
    
     \centering
     \begin{subfigure}[t]{0.325\linewidth}
         \centering
         \includegraphics[width=\textwidth]{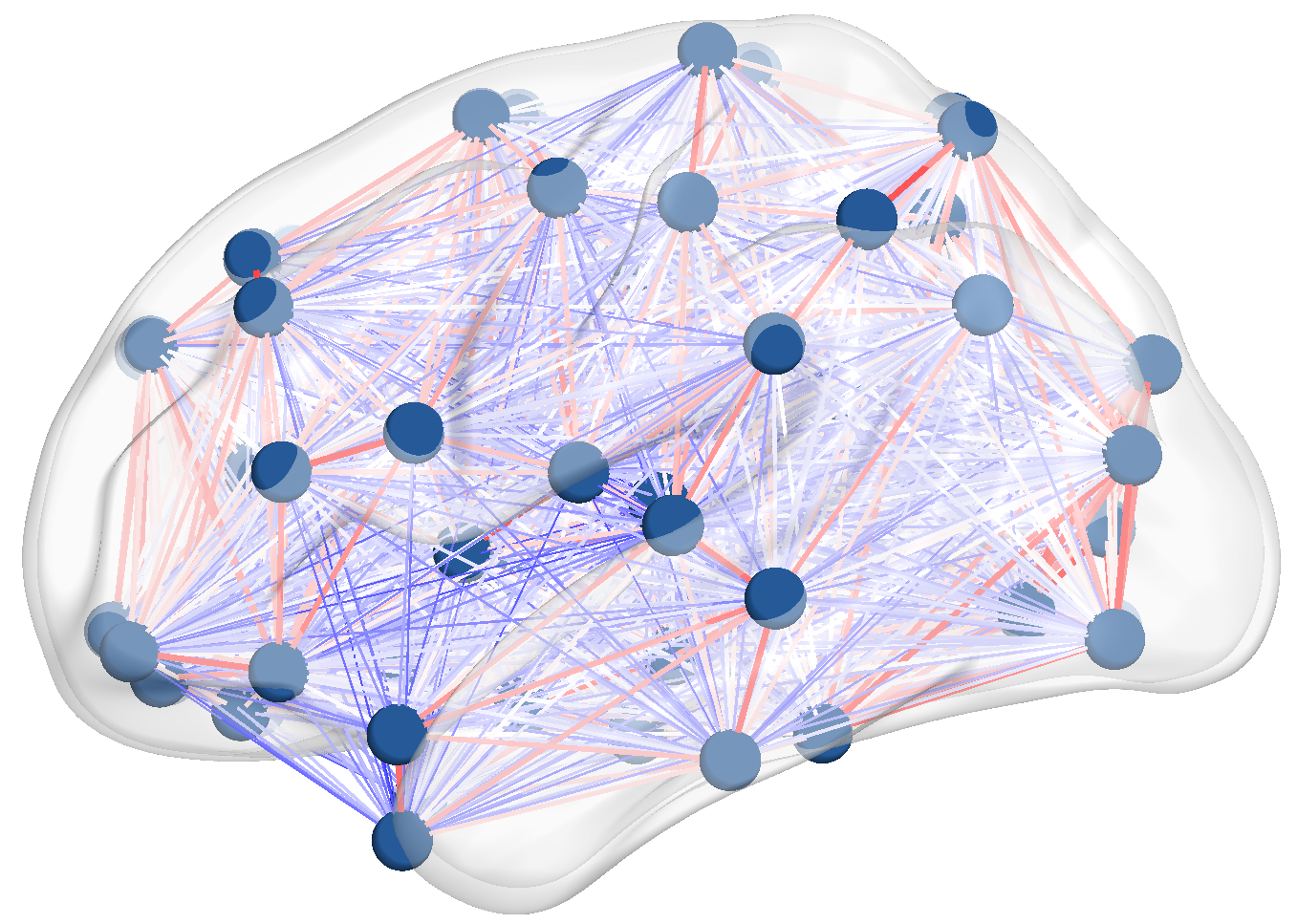}
     \end{subfigure}
     \hfill
     \begin{subfigure}[t]{0.325\linewidth}
         \centering
         \includegraphics[width=0.75\textwidth,height=0.75\textwidth]{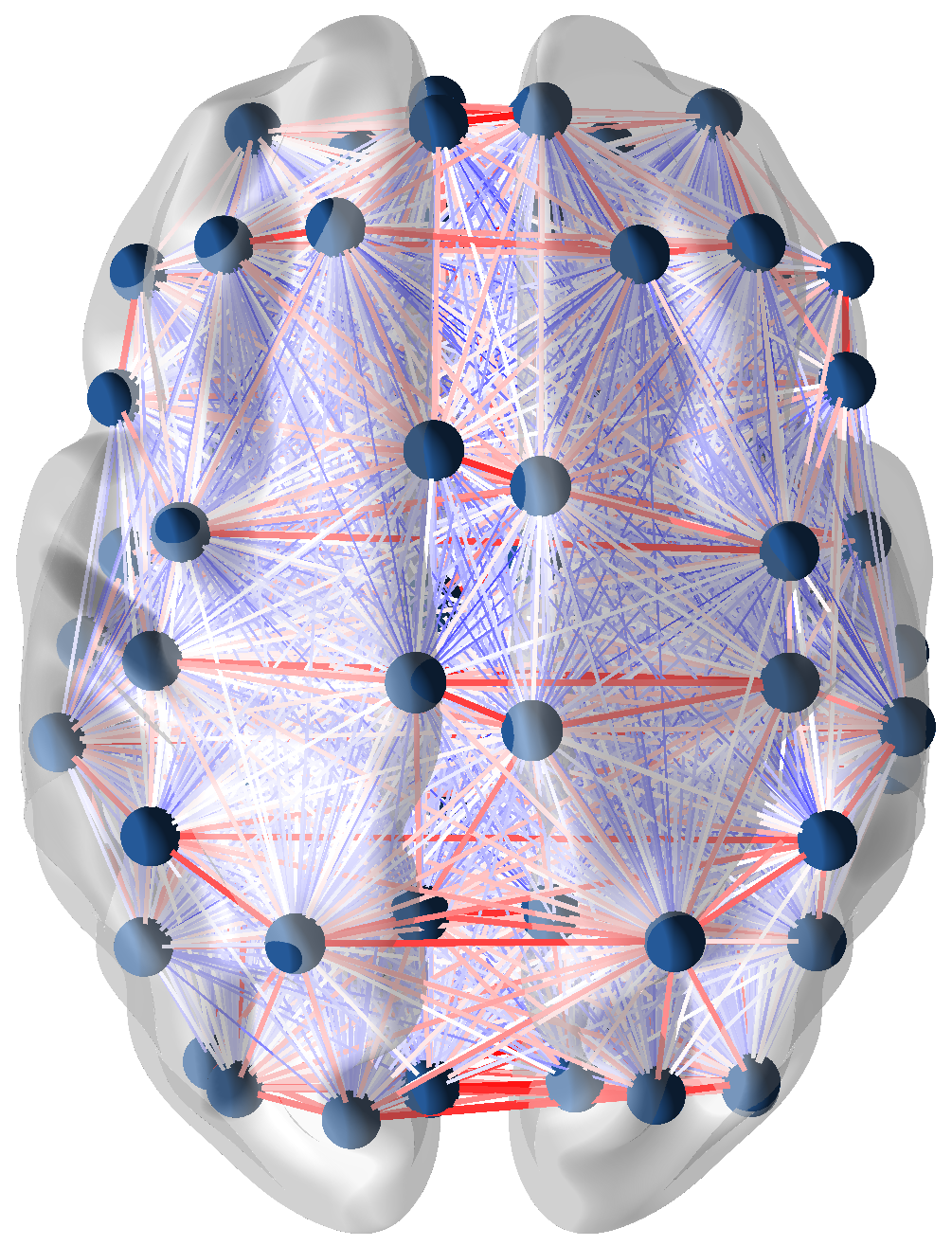}
     \end{subfigure}
     \hfill
     \begin{subfigure}[t]{0.325\linewidth}
         \centering
         \includegraphics[width=\textwidth]{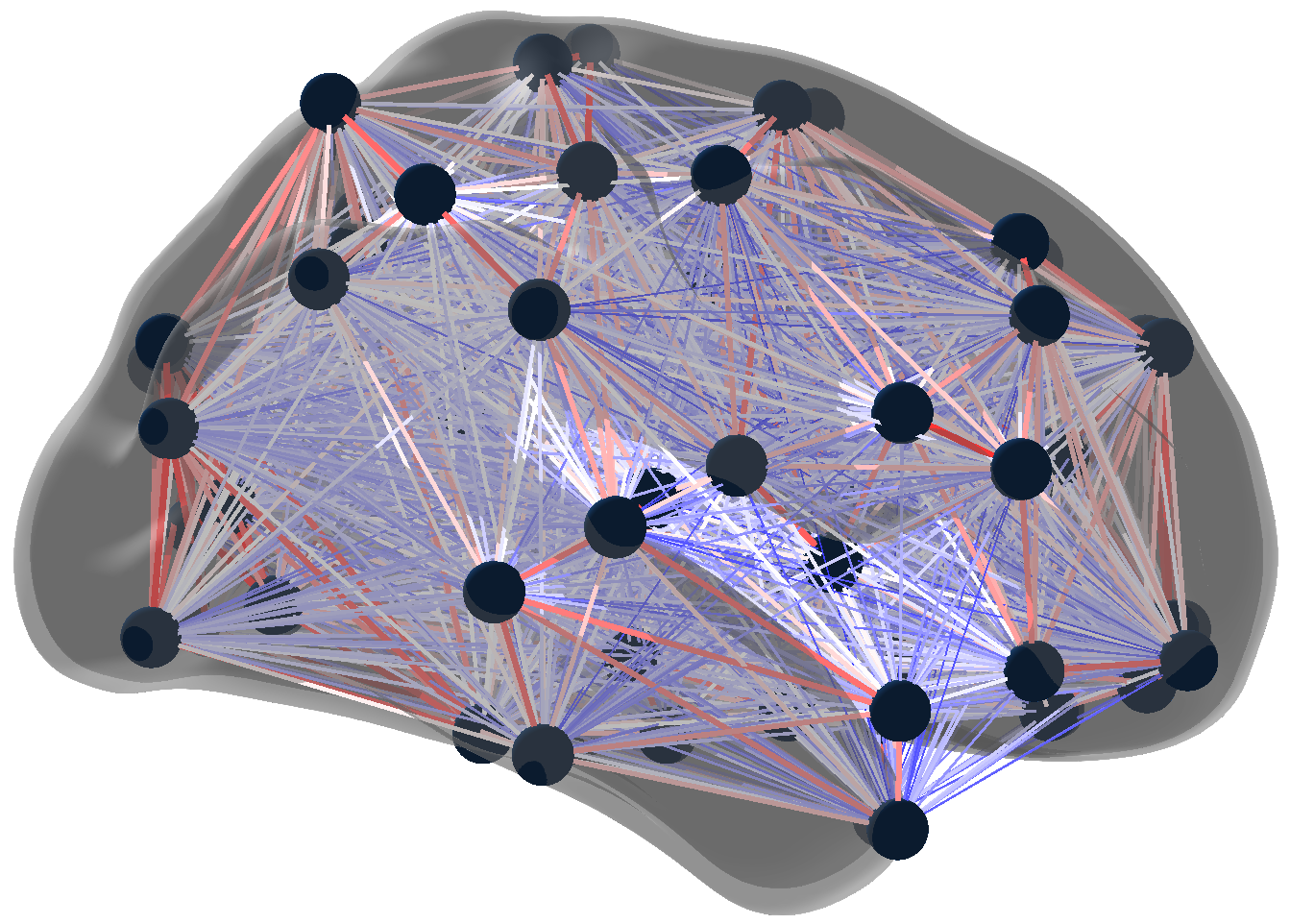}
     \end{subfigure}
      \begin{subfigure}[t]{1\linewidth}
         \centering
         \includegraphics[width=0.75\textwidth]{cb_use1.png}
     \end{subfigure}
    \caption{Top: The Loop Flow of non-users. Relatively few negative connections were observed. Bottom: The Loop Flow of users. Relatively few negative connections were observed. Notice it exhibits strong interhermispheric connections across networks compared to the Non-Loop Flow networks.}
    \label{fig:curl-flow}
\end{figure*}

Detailed spatial distributions of the decomposed edge flows are shown in Figures~\ref{fig:grad-flow} and~\ref{fig:curl-flow}, comparing user and non-user groups. The Loop Flow (Figure~\ref{fig:curl-flow}) retains fine-grained modular and cross-hemispheric organization, highlighting both intra- and interhemispheric cyclic connectivity patterns. It also emphasizes interhemispheric connections near the longitudinal fissure, highlighting long-range integrative pathways. In contrast, the Non-Loop Flow (Figure~\ref{fig:grad-flow}) exhibits smooth large-scale gradients dominated by global potential-driven structure, reflecting broad integrative processes rather than localized circuit interactions. Specifically, it captures predominantly intrahemispheric short-range associations, reflecting local circuit organization.

\begin{figure*}[ht]
     \centering
     \includegraphics[width=\textwidth, height=0.69\textwidth]{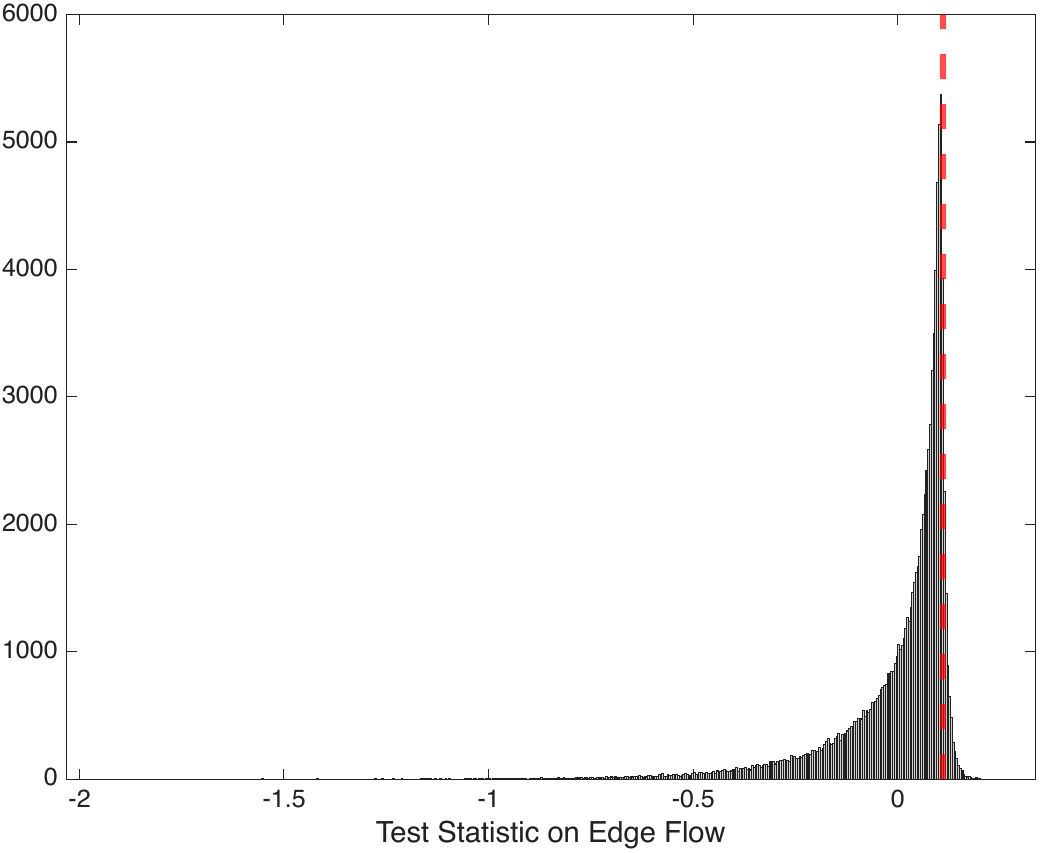}
    \caption{The distribution of the test statistic on the Edge Flow. The p-value obtained is 0.0771.}
    \label{fig:test-stats-edge}
\end{figure*}

\begin{figure*}[ht]
     \centering
     \includegraphics[width=\textwidth]{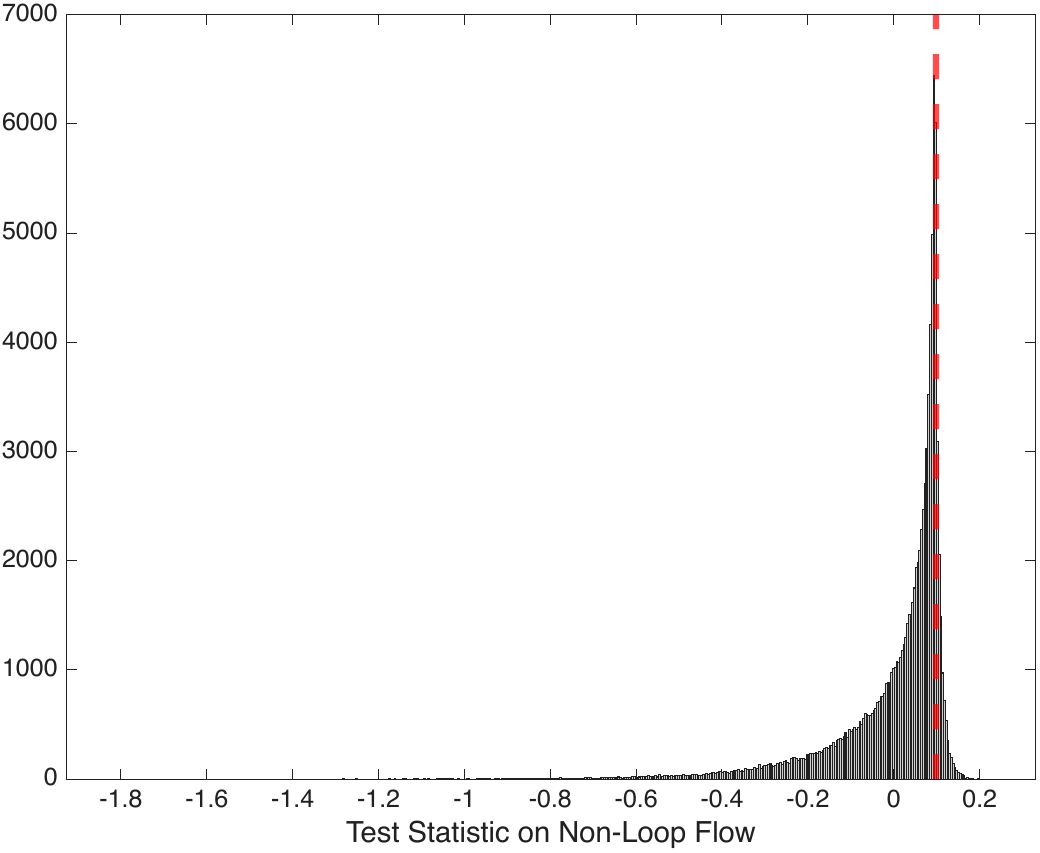}
    \caption{The distribution of the Non-Loop Flow test statistic. The p-value obtained is 0.1210.}
    \label{fig:test-stats-non-loop}
\end{figure*}

\begin{figure*}[ht]
     \centering
     \includegraphics[width=\textwidth, height=0.69\textwidth]{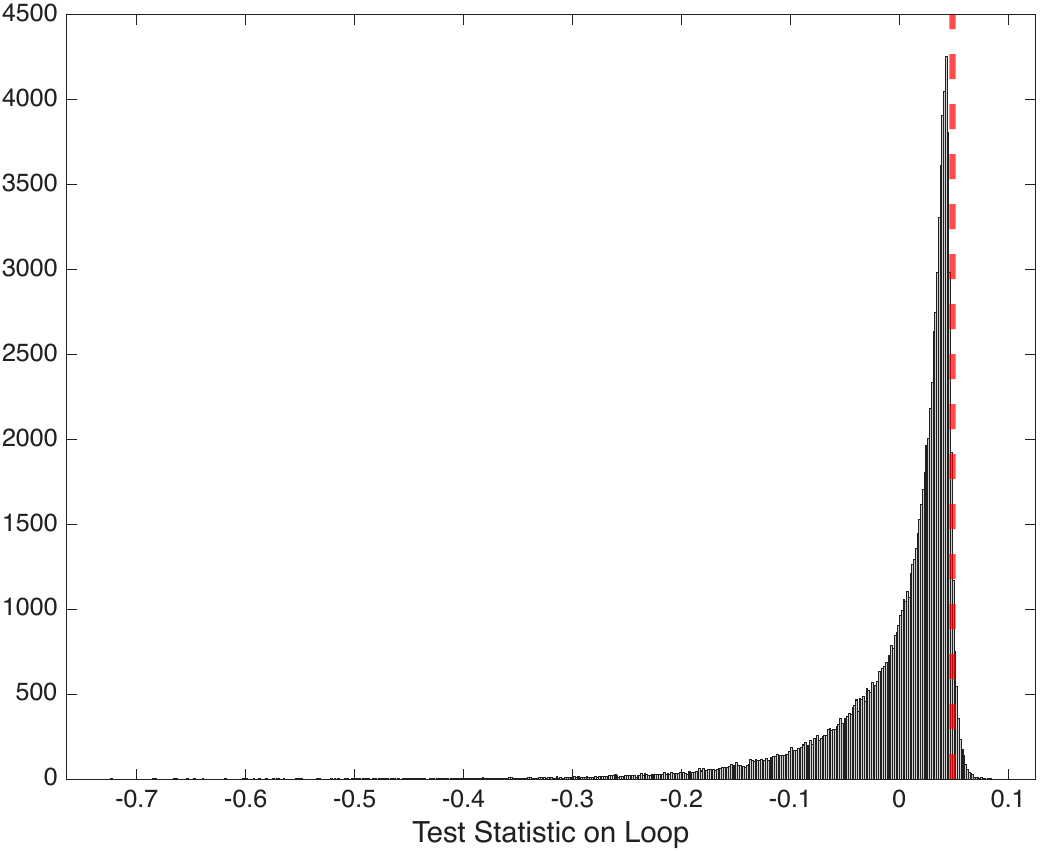}
    \caption{The distribution of the Loop Flow test statistic. The p-value obtained is 0.0370.}
    \label{fig:test-stats-loop}
\end{figure*}

The global test of topological equivalence using $100,000$ permutations is performed for each flow component. Figures~\ref{fig:test-stats-edge} - ~\ref{fig:test-stats-loop} displays the null distributions of the test statistic $\mathcal{T}$ along with the observed values (indicated by vertical red dashed-line).

For the Edge Flow, the obtained $p$-value is $0.0771$. This result suggests marginal evidence of topological differences in the raw connectivity structure, potentially driven by the Loop Flow component of the networks. Specifically, the Loop Flow comparison produced a $p$-value of $0.0370$, providing statistically significant evidence of topological differences between groups. This finding indicates that the global cyclical structures captured by the Loop Flow differs between cannabis users and non-users beyond what can be attributed to random variation. The Non-Loop Flow comparison yielded a $p$-value of $0.1210$, indicating no significant topological difference in the gradient component between groups. This supports the interpretation that potential-driven connectivity patterns remain largely preserved across groups, while global cyclic structures exhibit measurable alterations.


The divergent outcomes across flow components reveal the value of the Hodge decomposition. While raw edge weights show marginal differences that could reflect noise amplification, and Non-Loop flows show no difference. The Loop Flow isolates a specific aspect of network topology, global circular flow patterns, where cannabis use is associated with measurable structural changes. This selective sensitivity demonstrates how decomposing networks into orthogonal flow components can localize topological variability to specific architectural features, providing more interpretable and potentially more meaningful comparisons than aggregate connectivity measures.

These results suggest that cannabis exposure may preferentially not affect local circuit organization but rather global network integration, a hypothesis that would be difficult to formulate or test using conventional connectivity analysis methods

\section{Discussion and Conclusion}
\label{sec:disc}

This work developed a rigorous statistical and combinatorial framework for comparing weighted networks by combining combinatorial Hodge theory with Wasserstein variance minimization. The approach addresses a fundamental challenge in network neuroscience and related fields: distinguishing meaningful topological differences from noise-driven perturbations when comparing populations of networks. The key innovation of the method lies in decomposing network structure into orthogonal flow components before quantifying topological variability. The Hodge decomposition separates edge weights into Non-Loop Flows (representing potential-driven processes), and Loop Flows (representing global cycles). This decomposition isolates functionally distinct aspects of network architecture that exhibit different sensitivities to perturbations. The simulation studies demonstrate that variance minimization in the Wasserstein space is most effective when applied to the loop flow component, which filters out local random fluctuations while preserving genuine structural differences.

Simulation results provide compelling evidence for the method's robustness. When two network populations differ only in the magnitude of random perturbations, a scenario that mimics measurement noise or biological variability, raw edge-based comparisons increasingly detect spurious differences as sample size grows. In contrast, Non-Loop Flow comparisons maintain appropriate specificity even at larger sample sizes, correctly attributing the observed variance to noise rather than structure. This behavior is critical for avoiding false discoveries in high-dimensional network data where multiple testing considerations already pose significant challenges.
 
Application to functional brain networks from the ACPI-MTA dataset reveals the interpretive advantages of this framework. While aggregate connectivity shows marginal differences between cannabis users and non-users, decomposing the networks isolates the topological variation to the Loop flow component. Such localization of effects would be difficult to achieve using conventional connectivity metrics or graph-theoretic measures that conflate multiple architectural features. The framework also provides a foundation for edge-level localization of differences through Theorem~\ref{them:initial-test}, enabling researchers to identify specific connections driving group differences while controlling family-wise error rates. This capability bridges the gap between global topological summaries and detailed connectivity patterns, offering both interpretability and statistical rigor.

In conclusion, this work demonstrates that combining Hodge decomposition with Wasserstein variance minimization provides a principled approach to robust network comparison. By decomposing structure before quantifying variability, the method achieves both noise suppression and interpretive clarity. Our validation through simulations and application to functional brain networks establishes this framework as a valuable tool for detecting and localizing topological differences in weighted network populations while maintaining appropriate statistical control.

\bibliographystyle{plain}
\bibliography{references}
\end{document}